\documentclass[12pt]{article}

\usepackage{amsmath}
\usepackage{amssymb}
\usepackage{graphicx}
\usepackage{cite}
\usepackage{color}
\usepackage{setspace}
\usepackage{caption}
\usepackage{authblk}
\usepackage{subfig}

\makeatletter
\renewcommand{\@biblabel}[1]{\quad#1.}
\makeatother
\newcommand{\eq}[2]{\begin{equation} #1 \label{#2}\end{equation}}
\newcommand{\eqarray}[2]{\begin{eqnarray} #1 \label{#2}\end{eqnarray}}
\newcommand{\kav}{\langle k\rangle}
\newcommand{\dif}{\mathrm{d}}

\begin{document}

\title{Hierarchy in directed random networks}
\date{}
\author[1,*]{Enys Mones}

\affil[1]{\small{Department of Biological Physics, E\"{o}tv\"{o}s Lor\'{a}nd University, P\'{a}zm\'{a}ny P\'{e}ter stny. 1/A, H-1117 Budapest, Hungary}}

\pagestyle{empty}

\maketitle
\thispagestyle{empty}

\let\oldthefootnote\thefootnote
\renewcommand{\thefootnote}{\fnsymbol{footnote}}
\footnotetext[1]{Corresponding author: \texttt{enys@hal.elte.hu}}
\let\thefootnote\oldthefootnote

\begin{abstract}
In recent years, the theory and application of complex networks have been quickly developing in a markable way due to the increasing amount of data from real systems and to the fruitful application of powerful methods used in statistical physics. Many important characteristics of social or biological systems can be described by the study of their underlying structure of interactions. Hierarchy is one of these features that can be formulated in the language of networks. In this paper we present some (qualitative) analytic results on the hierarchical properties of random network models with zero correlations and also investigate, mainly numerically, the effects of different type of correlations. The behavior of hierarchy is different in the absence and the presence of the giant components. We show that the hierarchical structure can be drastically different if there are one-point correlations in the network. We also show numerical results suggesting that hierarchy does not change monotonously with the correlations and there is an optimal level of non-zero correlations maximizing the level of hierarchy.
\end{abstract}

\maketitle

\section{Introduction}
The application of complex networks in a broad range of social and biological systems has been the subject of much interest recently \cite{castellano09, dunne02, jeong00, negyessy06, cancho01}. These applications involve -- among others -- the description of ``small-world'' property of the networks \cite{watts98} or the consequences of scale-free degree distributions \cite{barabasi02}. Many aspects of the real systems can be studied in the framework of undirected networks \cite{newman10,newman03,pastorsatorras01,pastorsatorras04}. The most important property of the network is the degree distribution $p_k$ which is the probability of a randomly chosen node having $k$ edges  \cite{barabasi02}. Other features of the network can be understood through the degree distribution (average length of the shortest path between nodes, average number of edges between a node's neighbors \cite{newman01}, the characteristics of epidemics on the network \cite{pastorsatorras01} or robustness against failures and attacks \cite{barabasi00}).

\begin{figure}[!h]
    \centering
	\mbox{
	\subfloat[\textbf{Undirected graph}]{\includegraphics[width=8cm]{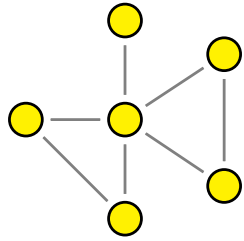}}
	\subfloat[\textbf{Directed graph}]{\includegraphics[width=8cm]{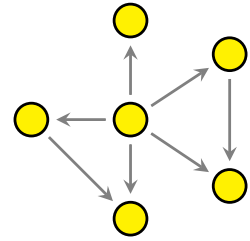}}}
	\mbox{
	\subfloat[\textbf{Giant component}]{\includegraphics[width=8cm]{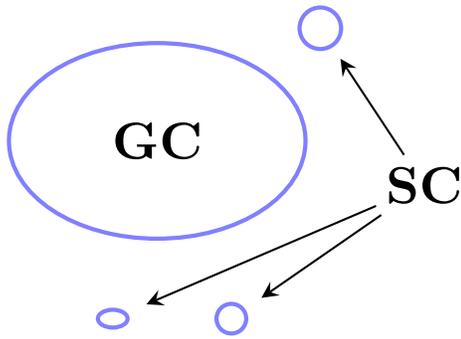}}
	\subfloat[\textbf{Giant components}]{\includegraphics[width=8cm]{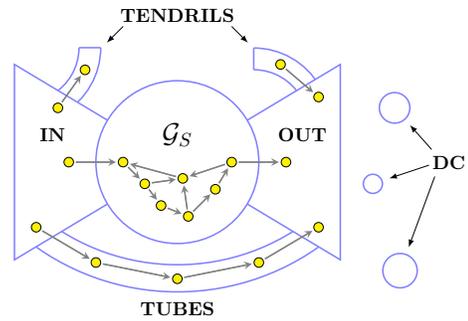}}}
	\caption{The difference between undirected (a) and directed (b) graphs. On large scale, the structure of the undirected graph (c) can be described by the largest connected component (the \emph{giant component} or GC) and the \emph{small components} (SC) \cite{bollobas01}. In the case of a directed graph (d), the giant components are best summarized in the ``bow-tie'' diagram \cite{broder00,dorogovtsev01}: the primary core is the \emph{giant strongly connected component} ($\mathcal{G}_{S}$). Inside the $\mathcal{G}_{S}$, every node can reach every other. There are nodes that can reach the whole $\mathcal{G}_{S}$ but not vice verse (IN) and together with the $\mathcal{G}_{S}$ they form the \emph{giant in component} ($\mathcal{G}_{in}$). The similar is true for the \emph{giant out component} ($\mathcal{G}_{out}$) but in the reversed direction. There are nodes that connect the $\mathcal{G}_{in}$ and $\mathcal{G}_{out}$ components but are not in the $\mathcal{G}_{S}$, they form the \emph{tubes}. There are also \emph{tendrils} that attach to the $\mathcal{G}_{in}$ and $\mathcal{G}_{out}$ components. The rest of the nodes are in the small \emph{disconnected components} (DC).}
	\label{fig:directed_network}
\end{figure}

However, most of the real networks are directed, i.e., the connection between two units of the system is not symmetric. Many structural properties of a directed network can be derived from undirected networks in a straightforward way \cite{newman10}, but the appearance of directionality also opens the door to features that are essentially different from those in undirected graphs (Fig.~\ref{fig:directed_network}).

\begin{figure}
    \centering
	\mbox{
	\subfloat[\textbf{Order hierarchy}]{\includegraphics[width=10cm]{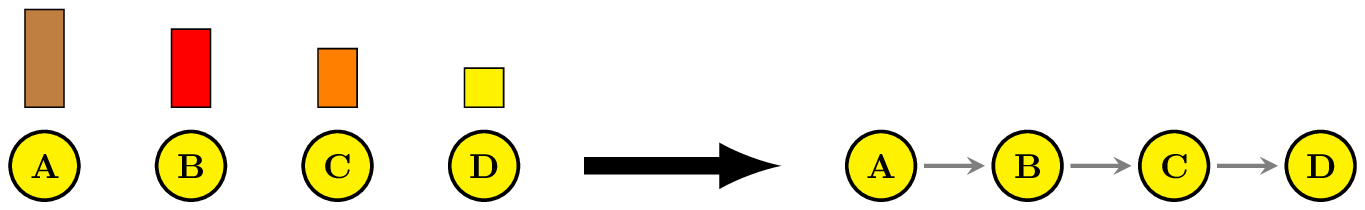}}}
	\mbox{
	\subfloat[\textbf{Nested hierarchy}]{\includegraphics[width=10cm]{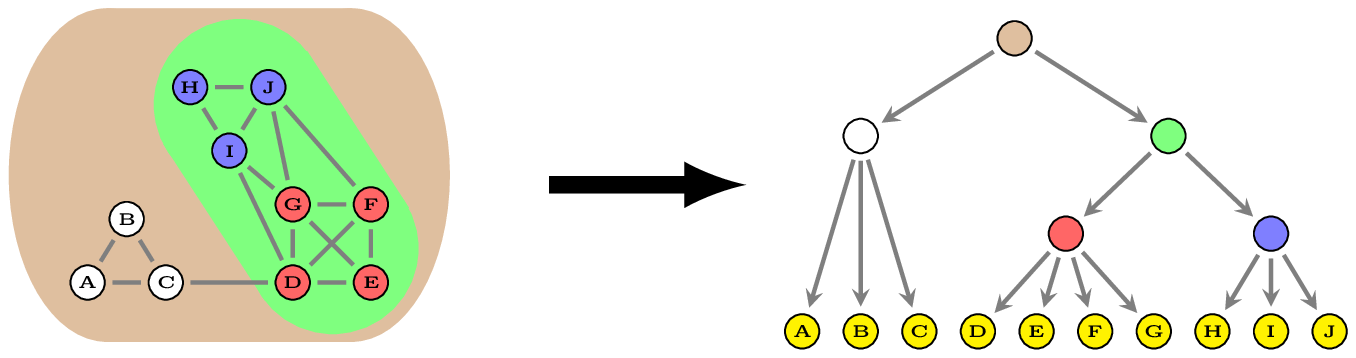}}}
	\mbox{
	\subfloat[\textbf{Flow hierarchy}]{\includegraphics[width=10cm]{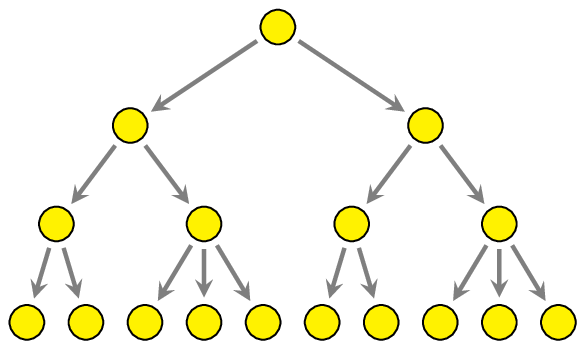}}}
	\caption{The three different types of hierarchy. (a) \emph{Order hierarchy} is simply a rank assigned to each unit making them an ordered set. (b) \emph{Nested hierarchy} is the hierarchy of the nested clusters of nodes. (c) In a (complete) \emph{flow hierarchy}, nodes can be layered in different levels so that the nodes that are influenced by other nodes (via an out-edge) are at lower levels. As the figures illustrate, each type can be transformed into a flow hierarchy. In the order hierarchy one can introduce a directed edge between every pair of adjacent nodes in the hierarchy. In the nested hierarchy one can assign a virtual node to each cluster and link a cluster to its contained clusters.}
	\label{fig:hierarchy_types}
\end{figure}

In the presence of directed edges, the organization level of the nodes on a large scale can be very complex and \emph{flow hierarchy} can emerge. It is a global structure of the network that is the result of the different roles of the nodes (see Fig.~\ref{fig:hierarchy_types} for the comparison of different hierarchy types).

As the growing number of findings show, hierarchy is a frequently appearing property of real networks, especially of networks describing social interactions \cite{nagy10,fushing11,ma04,huseyn84,pumain06}. The concept of hierarchy has led to different definitions (Fig.~\ref{fig:hierarchy_types}) and also to algorithms for measuring it in both directed and undirected networks \cite{lane06,wimberley09,carmel02,trusina04,rowe07,memon08}. In this paper we investigate \emph{flow hierarchy} by a corresponding measure, the \emph{global reaching centrality} ($G_R$), that has the intention to quantify that \cite{mones12}. In the following sections we show that the behavior of the $G_R$ is different below and above the critical average degree $k_c$, i.e., in the absence and presence of the giant components. We also give an approximation to its dependence on the average degree for different random network models when there are no degree correlations. In the last section we study the effects of degree correlations.

\section{Reaching centralities in uncorrelated random networks}
\subsection{Local and global reaching centrality}
Given a directed graph $G(V,E)$ with $N=|V|$ vertices and $M=|E|$ edges, the \emph{local reaching centrality} of node $i$ is defined as the number of reachable nodes via out-degrees divided by the total number of nodes:
\eq{c_R(i)=\frac{|S_i|}{N-1}=\frac{C_R(i)}{N-1}}{eq:cR_def}
where $S_i=\{j\in V|0<d^{out}(i,j)<\infty\}$ is the set of nodes that has finite, non-zero out-distance from node $i$. We will denote the size of the reachable set (i.e., the local reaching centrality without normalization) by $C_R$. The global reaching centrality of the graph is the normalized sum of the distances from the maximum local reaching centrality:
\eq{G_R=\frac{1}{N-1}\sum_i\Big[c_R^{max}-c_R(i)\Big]}{eq:grc_def1}
The normalization factor is the maximum possible value of the sum in a graph with $N$ nodes (this can be achieved in a star graph). The definition of the $G_R$ can be written in a more expressive form:
\eq{G_R=\frac{N}{(N-1)^2}\Big[C_R^{max}-\langle C_R\rangle\Big]}{eq:grc_def2}
Hence, the $G_R$ is proportional to the difference between the average and the maximum size of the reachable sets.

\subsection{Generating function method in directed networks}
The calculation of the reachable set of a node is equivalent to the problem of finding the out-component, which is the union of the reachable set and the node itself. The out-component can be determined by the generalization of the generating function formalism developed by Newman et al. \cite{newman01, newman10} to directed networks. Assuming that our graph has joint degree distribution $p_{ij}$, that is the probability of a randomly chosen node having $i$ in-degree and $j$ out-degree, the corresponding double generating function can be defined as
\eq{g_{00}(x,y)=\sum_{i,j=0}^\infty p_{ij}x^iy^j}{eq:g00_def}
and the generating functions for the excess in- and out-degree distributions \cite{newman10}:
\eq{g_{10}(x,y)=\frac{1}{\kav}\partial_xg_{00}(x,y)}{eq:g10_def}
\eq{g_{01}(x,y)=\frac{1}{\kav}\partial_yg_{00}(x,y)}{eq:g01_def}
Let $\pi_s^{out}$ denote the probability that a randomly chosen node has out-component of size $s$ and $\rho_s^{out}$ the probability that a randomly chosen edge points to an out-component of size $s$. Their generating functions are:
\eq{h_0(y)=\sum_{s=1}^\infty \pi_s^{out}y^s}{eq:h0_def}
\eq{h_1(y)=\sum_{s=1}^\infty \rho_s^{out}y^s}{eq:h1_def}
If we assume that the graph is \emph{locally tree-like} (i.e., loops are infrequent), they satisfy the following equations \cite{newman10}:
\eq{h_0(y)=yg_{00}[1,h_1(y)]}{eq:h0_eq}
\eq{h_1(y)=yg_{10}[1,h_1(y)]}{eq:h1_eq}
Using these equations for the generating functions, it is possible to derive a closed formula for the out-components in directed graphs (for the details see Appendix A):
\eq{\pi_s^{out}=\frac{\kav}{(s-1)!}\Bigg[\frac{\dif^{s-2}}{\dif y^{s-2}}\bigg(g_{01}(1,y)g_{10}(1,y)^{s-1}\bigg)\Bigg]_{y=0}}{eq:pi_formula}
Thus, if we know the joint degree distribution, we can calculate $P(C_R)$ (and thus the distribution of the local reaching centralities as well, since it differs from $P(C_R)$ only by a scale factor). In order to calculate $P(C_R)$, we have to determine the out-components and apply the $s=C_R+1$ substitution. With the knowledge about $P(C_R)$, we are able to calculate the $G_R$ at different average degrees.

It is important to note that Eq.~(\ref{eq:pi_formula}) describes only the \emph{small} reaching centralities, i.e., the distribution of local reaching centralities of the nodes outside the giant components. Above the critical average degree, where the giant components are already present, a significant fraction of the nodes can reach finite fraction of the whole network. In this regime, there are three type of nodes besides the ones with small reaching centrality: the nodes inside the $\mathcal{G}_{S}$, $\mathcal{G}_{in}$ and $\mathcal{G}_{out}$, each type can reach differenct fraction of the graph. Thus, in the distribution of the local reaching centrality above the percolation threshold, a peak with finite width appears at large values.

Before applying these equations to random networks with different degree distributions, we determine the $G_R$ for the case of a hierarchical tree.

\subsection{Hierarchical tree}
\subsubsection{Local reaching centralities}
\begin{figure}
    \centering
	\includegraphics[width=10cm]{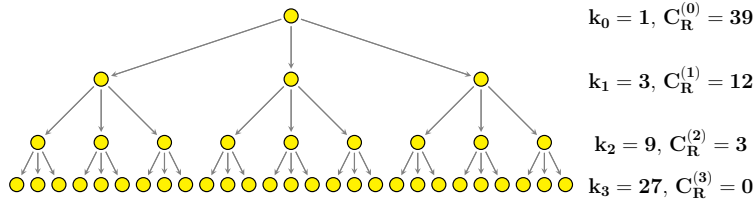}
	\caption{The number of nodes and size of reachable sets in the different levels.}
	\label{fig:tree_levels}
\end{figure}

\begin{figure}
    \centering
	\includegraphics[angle=-90,width=10cm]{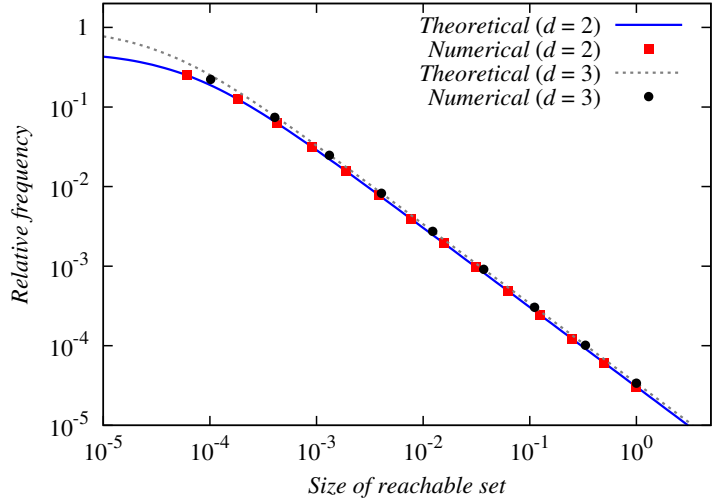}
	\caption{Theoretical and measured distribution of the local reaching centrality in the hierarchical tree. The dots are the results from graphs with $d=2$ (red squares) and $d=3$ (black circles), the corresponding number of nodes are $N_2=32767$ and $N_3=29524$.}
	\label{fig:cr_tree}
\end{figure}

In a tree graph with $N$ vertices and branching number of $d$, nodes in the same level have the same local reaching centrality. Let us denote the size of reachable set of a node in the $\ell$-th level by $C_R^{(\ell)}$. The number of nodes and the size of the reachable set in the $\ell$-th level are the following (see Fig.~\ref{fig:tree_levels}):
\eq{k_\ell=d^\ell}{eq:tree_kl}
\eq{C_R^{(\ell)}=\frac{N-\sum_{j=0}^\ell d^j}{d^\ell}=\frac{N-\frac{d^{\ell+1}-1}{d-1}}{d^\ell}}{eq:tree_CRl}
The probability that a randomly chosen node has a reachable set of size $C_R^{(\ell)}$ is $k_\ell/N$. Substituting Eq.~(\ref{eq:tree_kl}) in Eq.~(\ref{eq:tree_CRl}) gives:
\eq{P\big(C_R^{(\ell)}\big)=\frac{N(d-1)+1}{N(d-1)C_R^{(\ell)}+Nd}}{eq:tree_PCR}
In the asymptotic limit of $N\to\infty$, we get the following approximation for the probability of the normalized local reaching centrality (we also omit the $\ell$ index, since in the mentioned limit, the allowed discrete values of $C_R^{(\ell)}/(N-1)$ are becoming dense in $[0;1]$ and $c_R^{(\ell)}$ becomes continuous) :
\eq{P(c_R)\approx\frac{1}{Nc_R+\frac{d}{d-1}}}{eq:tree_PcR}
Comparison with numerical results is plotted in Fig.~\ref{fig:cr_tree}.
In the simulations, trees with branching number of 2 and 3 are used, the corresponding number of levels are 15 and 10. The numerical results show satisfactory agreement with Eq.~(\ref{eq:tree_PcR}).

\subsubsection{Global reaching centrality}
\begin{figure}
    \centering
	\includegraphics[angle=-90,width=10cm]{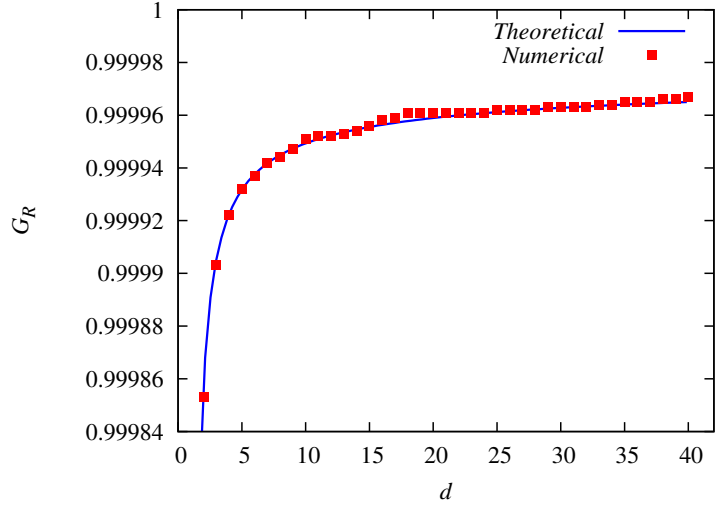}
	\caption{Comparison of the measured $G_R$ with the theoretical prediction. The dots are the $G_R$ of trees with $N=10^5$ nodes. Because of the fixed number of nodes, the number of levels are approximated by Eq.~(\ref{eq:tree_L_def_modified}), which neglects the lowermost nodes in the total number of nodes. The deviations from the theoretical curve depends on the branching number. For branching numbers with an integer power close to $10^5$, the theoretical curve is more accurate.}
	\label{fig:grc_tree}
\end{figure}

Using the equations for the number of nodes and size of reachable sets in the $\ell$-th level, we can easily calculate the size of the reachable sets:
\eq{\langle C_R\rangle=\sum_{\ell=0}^{L-1}P\big(C_R^{(\ell)}\big)\cdot C_R^{(\ell)}=\sum_{\ell=0}^{L-1}\bigg(1-\frac{d^{\ell+1}-1}{N(d-1)}\bigg)}{eq:tree_CR_av1}
where $L$ denotes the number of levels (the level of the root is zero). This number can be determined by the constraint that the sum of the nodes in all levels gives the number of nodes:
\eq{\sum_{\ell=0}^{L-1}d^\ell=\frac{d^L-1}{d-1}=N}{eq:tree_L_def}
Arranging for $L$ and simplifying it:
\eq{L=\frac{\ln[N(d-1)+1]}{\ln d}}{eq:tree_L_eq}
Calculating the sum in the right hand side of Eq.~(\ref{eq:tree_CR_av1}) gives:
\eq{\langle C_R\rangle=L\bigg(1+\frac{1}{N(d-1)}\bigg)-\frac{d}{(d-1)}}{eq:tree_CR_av2}
Substituting $L$ into the last equation and using Eq.~(\ref{eq:grc_def2}) we get:
\begin{equation}
    G_R=\frac{N}{(N-1)^2}\Bigg[\underbrace{N-1}_{C_R^{max}}-\underbrace{\frac{\ln[N(d-1)+1]}{\ln d}\bigg(1+\frac{1}{N(d-1)}\bigg)+\frac{d}{(d-1)}}_{\langle C_R\rangle}\Bigg]
    \label{eq:tree_grc}
\end{equation}
In the asymptotic case of $N\to\infty$:
\eq{G_R=1+\frac{2d-1}{N(d-1)}-\frac{\ln[N(d-1)+1]}{N\ln d}}{eq:tree_grc_asymptotic}
In Fig.~\ref{fig:grc_tree} we show the comparison of Eq.~(\ref{eq:tree_grc}) with the simulations.
In the numerical results, the number of nodes is $10^5$ for every branching number. Thus, Eq.~(\ref{eq:tree_L_def}) for the number of levels does not hold exactly: there are less nodes in the last level than expected. This fact is taken into account by summing up only to $L-2$ in Eq.~(\ref{eq:tree_L_def}):
\eq{\sum_{\ell=0}^{L-2}d^\ell=N}{eq:tree_L_def_modified}
This means that in the expression of $L$, the number of nodes is replaced by a reduced number (which is the nodes at the lowermost level). This modification gives a better agreement with the numerical calculations for large branching numbers. From Eq.~(\ref{eq:tree_L_def}), it is also clear that the approximation is more accurate if a branching number has an integer power close to $10^5$. This can be observed in Fig.~\ref{fig:grc_tree} as well.

\subsection{Erd\H{o}s--R\'{e}nyi graph}
\subsubsection{Local reaching centralities}
\begin{figure}
    \centering
	\includegraphics[angle=-90,width=10cm]{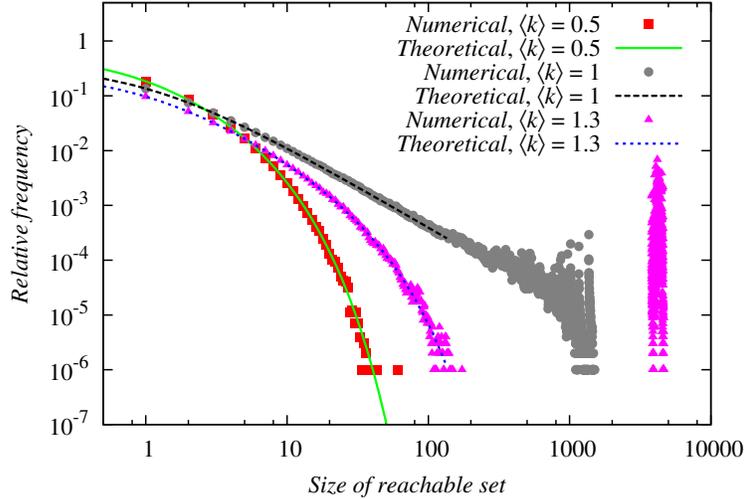}
	\caption{Distribution of the small reachable set sizes for the ER graph, based on the assumption that the graph is locally tree-like. Below the transition point (red squares), there is no node that can reach any finite part of the graph. This changes when the $\mathcal{G}_{S}$ appears at $\langle k\rangle=1$ (gray circles). The numerical results are the averages of 1000 independent calculations on networks with $N=10^4$. Note that Eq.~(\ref{eq:ER_PCR}) describes the distribution of the small components, thus the emerging $\mathcal{G}_{S}$ (the peak at large reachable sets) is out of its scope.}
	\label{fig:cr_er}
\end{figure}

In the case of uncorrelated in- and out-degrees, the joint degree distribution of an Erd\H{o}s--R\'{e}nyi (ER) graph \cite{erdos60,bollobas01} is the product of two independent Poisson distributions:
\eq{p_{ij}=e^{-2\kav}\frac{\kav^{i+j}}{i!j!}}{eq:ER_pij}
where $\kav$ denotes the average degree. The double generating function has the form:
\eq{g_{00}(x,y)=e^{\kav(x+y-2)}}{eq:ER_g00}
which is also the generating function of the excess degree distributions in this case. Now, using Eq.~(\ref{eq:pi_formula}) we get:
\eqarray{\pi_s^{out}	& = & \frac{\kav}{(s-1)!}\Bigg[\frac{\dif^{s-2}}{\dif y^{s-2}}\bigg(e^{\kav s(y-1)}\bigg)\Bigg]_{y=0}= \nonumber \\
			& = & \frac{\kav^{s-1}s^{s-2}}{(s-1)!}e^{-\kav s}}{eq:ER_pi}
And for the size of small reachable sets (without the giant components):
\eq{P(C_R)=\frac{\kav^{C_R}(C_R+1)^{C_R-1}}{C_R!}e^{-\kav(C_R+1)}}{eq:ER_PCR}
Fig.~\ref{fig:cr_er} shows this result compared with the numerical distributions.
It is important to understand the limits of Eq.~(\ref{eq:ER_PCR}). It is only the distribution for the size of the \emph{small reachable sets}, i.e., it does not contain the $\mathcal{G}_{S}$\footnote{However, the integrated size of the giant components is encoded in the distribution, since it is normalized to \mbox{$N-|\mathcal{G}_{W}|$}, where $\mathcal{G}_{W}$ is the \emph{giant weakly connected component}.}. This can be seen on the plots. Before the transition, all of the nodes are in separated small components and the reaching centrality distribution vanishes very quickly. Above the transition point, they start to aggregate in the $\mathcal{G}_{S}$. Note that in this regime, where the giant component appears, a large amount of nodes have a large reachable set. More precisely, the nodes in the $\mathcal{G}_{in}$ component can reach every node in $\mathcal{G}_{out}$. This is clearly seen in Fig.~\ref{fig:cr_er}: the peak has a finite width, corresponding to the three different giant components.

\subsubsection{Global reaching centrality}
\begin{figure}
    \centering
	\includegraphics[angle=-90,width=10cm]{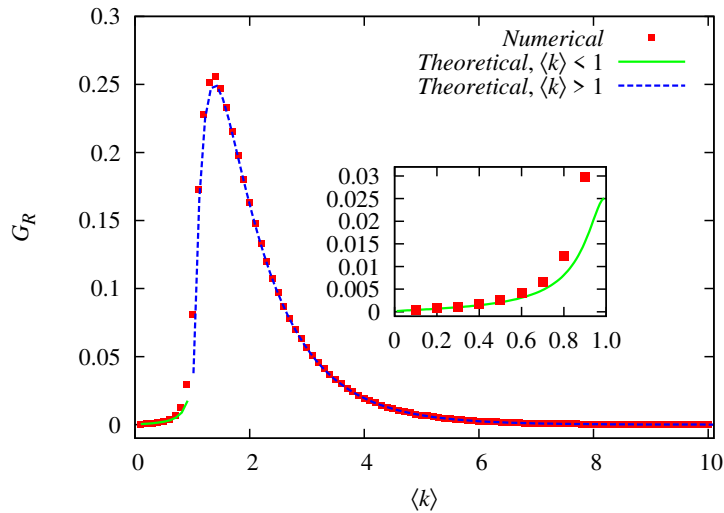}
	\caption{Global reaching centrality in the ER graph. The dots are the average of 1000 independent calculations on networks with $N=10^4$. The errors are comparable to the size of the dots. The lines show the different approximations in the two regimes: below the percolation threshold (solid green line) and above it (dashed blue line). Both approximations tend to deviate from the numerical values near the transition point.}
	\label{fig:grc_er}
\end{figure}

\begin{figure}
    \centering
	\includegraphics[angle=-90,width=10cm]{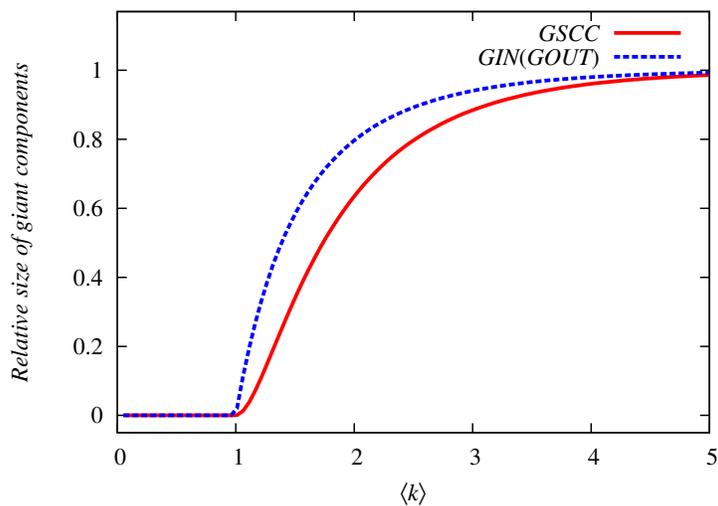}
	\caption{The relative sizes of the giant components of the bow-tie diagram in the ER network (Fig.~\ref{fig:directed_network}d). Although $\mathcal{G}_{out}$ (dashed blue line) grows quickly with the average degree, at low edge densities ($\kav<2$) both components contain less than 80\% of the nodes.}
	\label{fig:er_giant_components}
\end{figure}

We have to distinguish between the graph without the $\mathcal{G}_{S}$ and with the $\mathcal{G}_{S}$ \cite{newman10}. Below the transition point $k_c$, there are only small components. All of the nodes can reach only infinitesimal part in the graph, thus having an average reachable set size of order unit ($\langle C_R\rangle=\mathcal{O}(1)$). In this regime, the $G_R$ is dominated by the maximum value of the local reaching centrality. Since most of the nodes can reach very few other nodes and the distribution vanishes quickly, we can assume that the largest reachable set belongs to only one node (and that the corresponding out-component has the smallest relative frequency). Given a graph with $N$ nodes, this condition translates as $P(C_R)\approx1/N$ for $C_R=C_R^{max}$. Thus, finding the reachable set size that has only one realization can lead us to find the largest component. In the $C_R^{max}\gg1$ limit we can use the Stirling-formula to approximate $P(C_R)$:
\eq{P(C_R)\approx e^{1-\kav}\frac{\big(\kav e^{1-\kav}\big)^{C_R}}{\sqrt{2\pi}(C_R+1)^{3/2}}}{eq:ER_PCR_asymptotic}
Writing the $P(C_R)\approx1/N$ condition and rearranging it for $C_R$ we get the following exponential equation:
\eq{\Bigg(\frac{\sqrt{2\pi}}{Ne^{1-\kav}}\Bigg)^{2/3}(C_R+1)=\exp\bigg[\frac{2}{3}(\ln \kav+1-\kav)C_R\bigg]}{eq:ER_CRmax_eq}
This equation can be solved in terms of the \emph{Lambert W} function \cite{corless96}. The equation of the form
\eq{A(x-R)=e^{-Bx}}{eq:ER_exponential_equation}
has the solution of $x=R+\frac{1}{B}W\Big[\frac{Be^{-Br}}{A}\Big]$. Now we have
\eq{A=\Big(\frac{\sqrt{2\pi}}{Ne^{1-\kav}}\Big)^{2/3}}{eq:ER_A}
\eq{B=\frac{2}{3}(\ln \kav+1-\kav)}{eq:ER_B}
\eq{R=-1}{eq:ER_R}
so the final expression for the largest local reaching centrality:
\begin{equation}
	C_R^{max}=-\frac{W\bigg[\frac{2}{3}(\ln \kav+1-\kav)e^{-\frac{2}{3}(\ln \kav+1-\kav)}\Big(\frac{\sqrt{2\pi}}{Ne^{1-\kav}}\Big)^{-2/3}\bigg]}{\frac{2}{3}(\ln \kav+1-\kav)}-1
    \label{eq:ER_CRmax}
\end{equation}
Now we turn our attention to the case when the $\mathcal{G}_{S}$ is already present. In this case, for a good approximation, we can ignore those nodes that are not in the bow-tie (they are relevant only in the average local reaching centrality, but they have only an infinitesimal contribution). Thinking of the ``bow-tie'' picture, we can assume that there are some nodes that can reach the whole bow-tie. Thus, $C_R^{max}\approx|\mathcal{G}_{in}|+|\mathrm{\mathcal{G}_{out}}|-|\mathcal{G}_{S}|$. The average is slightly different and nontrivial, but let assume that it is equivalent to the size of the giant out-component $\mathcal{G}_{out}$. If we assume that most of the nodes are gathering in the $\mathcal{G}_{S}$, it is also reasonable that the average size of reachable sets is dominated by the nodes in the $\mathcal{G}_{S}$ and they can reach the whole $\mathcal{G}_{out}$ component (see Fig.~\ref{fig:directed_network}d). Using these assumptions, the size of the reachable sets is approximately $|\mathrm{\mathcal{G}_{out}}|$. The relative sizes of these components are quite the same as the local reaching centrality. In the generating function formalism they are given by:
\eq{\frac{|\mathcal{G}_{S}|}{N}=S=1-g_{00}(u,1)-g_{00}(1,v)+g_{00}(u,v)}{eq:ER_gscc_eq}
\eq{\frac{|\mathcal{G}_{in}|}{N}=I=1-g_{00}(u,1)}{eq:ER_gin_eq}
\eq{\frac{|\mathrm{\mathcal{G}_{out}}|}{N}=O=1-g_{00}(1,v)}{eq:ER_gout_eq}
where $u$ and $v$ are the smallest non-zero solutions of the following equations \cite{newman10}:
\eq{u=g_{01}(u,1)}{eq:ER_u_eq}
\eq{v=g_{10}(1,v)}{eq:ER_v_eq}
Since the generating function is symmetrical in its variables, we have $u=v$. Using Eq.~(\ref{eq:ER_g00}), the solution can be written in terms of the Lambert W function giving:
\eq{u(\kav)=-\frac{1}{\kav}W(-\kav e^{-\kav})}{eq:ER_u_result}
And for the $G_R$ above the phase transition:
\eq{G_R=I-S=g_{00}[u(\kav),1]-g_{00}[u(\kav),u(\kav)]}{eq:ER_grc}
The theoretical curve and the measurements are shown in Fig.~\ref{fig:grc_er}.
The theoretical curve in the $k_c<\kav<2$ range is a little below the numerical results. This is in good accordance with the assumptions used in deriving Eq.~(\ref{eq:ER_grc}): the average reachable set is approximated by the $\mathcal{G}_{out}$ component, however at small average degrees, there are many small out-components that decrease the average. The relative sizes of the giant components are depicted on Fig.~\ref{fig:er_giant_components}.
It is clearly seen that near the critical average degree, an appreciable portion of the nodes is outside of the giant components and they all have very small reachable set. Thus, the difference between the largest and average reaching centrality is larger in the real network, resulting in a larger $G_R$. The theoretical curve below $k_c$ predicts lower $G_R$, which indicates that the distribution of $P(C_R)$ described by Eq.~(\ref{eq:ER_PCR}) is no more valid for the whole network when approaching the critical average degree.

\subsection{Exponential network}
\subsubsection{Local reaching centralities}
\begin{figure}
    \centering
	\includegraphics[angle=-90,width=10cm]{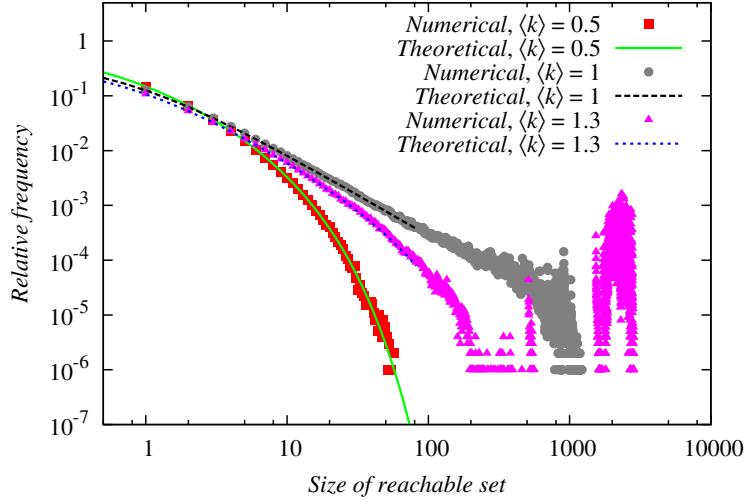}
	\caption{Distribution of the small reachable set sizes in the exponential graph. Numerical results are the average of 1000 measurement on networks with $N=10^4$. The emergence of the giant components (which are not included by the analytical distribution) can be clearly seen at $\kav=1$ (gray circles) and $\kav=1.3$ (purple triangles).}
	\label{fig:cr_ex}
\end{figure}

In this section we calculate $P(C_R)$ and the $G_R$ for the exponential network, motivated by the finding that the distribution of many real world networks can be well fitted by an exponential \cite{amaral00}. An uncorrelated exponential network has joint degree distribution of the form
\eq{p_{ij}=(1-e^{-1/\kappa})^2e^{-(i+j)/\kappa}}{eq:EXP_pij}
where the average degree can be obtained from the $\kappa$ parameter: $\kav=(e^{1/\kappa}-1)^{-1}$. The double generating function and its derivatives are:
\eq{g_{00}(x,y)	= \frac{(e^{1/\kappa}-1)^2}{(e^{1/\kappa}-x)(e^{1/\kappa}-y)}}{eq:EXP_g00}
\eq{g_{10}(x,y)=\frac{1}{\kav}\frac{(e^{1/\kappa}-1)^2}{(e^{1/\kappa}-x)^2(e^{1/\kappa}-y)}}{eq:EXP_g10}
\eq{g_{01}(x,y)=\frac{1}{\kav}\frac{(e^{1/\kappa}-1)^2}{(e^{1/\kappa}-x)(e^{1/\kappa}-y)^2}}{eq:EXP_g01}
By the formula for the out-components we have:
\eq{\pi_s^{out}	= \frac{1}{\kav^s(s-1)!}\frac{(2s-2)!}{s!}e^{-\frac{2s-1}{\kappa}}}{eq:EXP_pi1}
Substituting $e^{1/\kappa}=\frac{\kav+1}{\kav}$ we get:
\eq{\pi_s^{out}=\frac{(2s-2)!}{\kav^s(s-1)!s!}\bigg(\frac{\kav}{\kav+1}\bigg)^{2s-1}}{eq:EXP_pi2}
Translating this to the small reachable sets (see Fig.~\ref{fig:cr_ex} for comparison with numerical calculations):
\eq{P(C_R)=\frac{(2C_R)!}{\kav^{C_R+1}C_R!(C_R+1)!}\bigg(\frac{\kav}{\kav+1}\bigg)^{2C_R+1}}{eq:EXP_PCR}

\subsubsection{Global reaching centrality}
\begin{figure}
    \centering
	\includegraphics[angle=-90,width=10cm]{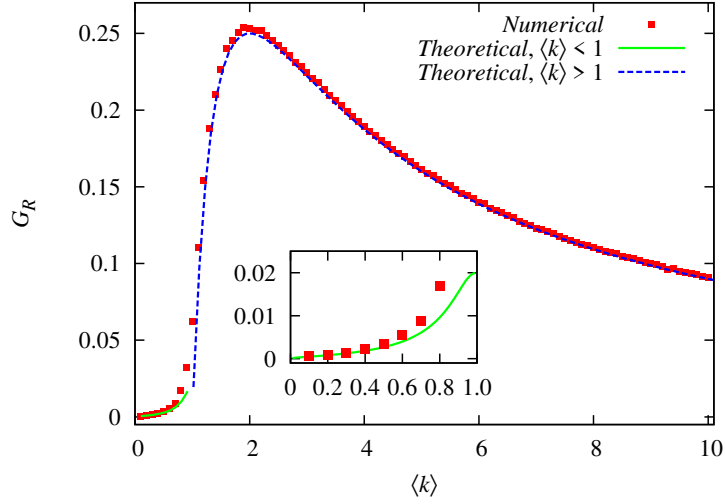}
	\caption{Global reaching centrality of the exponential graph and the predicted curves in the different regimes of $\kav$. Every numerical point is the average of 1000 independent runs on networks with $N=10^4$. The magnitude of the errors is in the order of the dot sizes.}
	\label{fig:grc_ex}
\end{figure}

The first task is to find the smallest solution of the $u=g_{01}(u,1)$ and $v=g_{10}(1,v)$ equations. Since we assume uncorrelated exponential distributions, we have $u=v$. The equation
\eq{u=g_{01}(u,1)=\frac{1}{\kav(e^{1/\kappa}-u)}}{eq:EXP_u_eq}
is quadratic in $u$ and has two solutions: $u_1=1$ and $u_2=1/\kav$, thus for $\kav<1$ there are no giant components and we can use the ``rarest component'' assumption as before with the ER graph. For this to do, we have to solve the following equation for $C_R$:
\eq{\frac{(2C_R)!}{\kav^{C_R+1}C_R!(C_R+1)!}\bigg(\frac{\kav}{\kav+1}\bigg)^{2C_R+1}=\frac{1}{N}}{eq:EXP_CRmax_eq1}
Using the Stirling-formula (without going into the details), this equation can be approximated by the following:
\begin{equation}
    \bigg(\frac{(\kav+1)\sqrt{\pi}}{Ne}\bigg)\bigg(\frac{C_R+1}{C_R}\bigg)^{C_R}(C_R+1)^{3/2}=e^{\frac{2}{3}\ln\Big(\frac{4\kav}{(\kav+1)^2}\Big)C_R}
    \label{eq:EXP_CRmax_eq2}
\end{equation}
We assume that the rarest component is much larger than one, thus $\big(\frac{C_R+1}{C_R}\big)^{C_R}\approx e$. This reduces our equations and we can rewrite it:
\eq{AC_R=e^{-BC_R}}{eq:EXP_exponential_equation}
Where we used the following shorthand notations:
\eq{A=\bigg(\frac{(\kav+1)\sqrt{\pi}}{N}\bigg)^{2/3}}{eq:EXP_A}
\eq{B=-\frac{2}{3}\ln\bigg(\frac{4\kav}{(\kav+1)^2}\bigg)}{eq:EXP_B}
And neglected the additional $1$ in the left hand side. The solution of Eq.~(\ref{eq:EXP_CRmax_eq2}) is then:
\eq{C_R^{max}=\frac{1}{B}W\bigg(\frac{B}{A}\bigg)}{eq:EXP_CRmax}
If $\kav>1$, this approximation fails because of the appearance of the giant components. Using the solution of Eq.~(\ref{eq:EXP_u_eq}), the relative sizes of the parts in the bow-tie diagram are:
\eq{S=\bigg(1-\frac{1}{\kav}\bigg)^2}{eq:EXP_S}
\eq{I=1-\frac{1}{\kav}}{eq:EXP_I}
\eq{O=1-\frac{1}{\kav}}{eq:EXP_O}
By the same argument as with the ER graph, we get for the $G_R$:
\eq{G_R=1-\frac{1}{\kav}-\bigg(1-\frac{1}{\kav}\bigg)^2}{eq:EXP_grc}
This result is shown on Fig.~\ref{fig:grc_ex} along with the numerical results.
The same argument can be applied for the exponential network as in the previous section. The predicted curves fit well in the very small and in the large average degree regimes.

\subsection{Scale-free network}
\begin{figure}
    \centering
	\includegraphics[angle=-90,width=10cm]{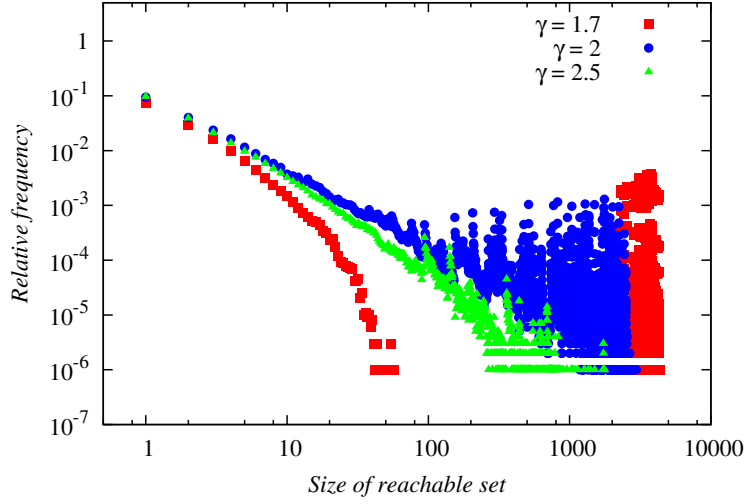}
	\caption{Distribution of the reachable set sizes in the scale-free graph. The data points are the average of 1000 distributions on networks with $N=10^4$. The giant components emerge at $\gamma=2$ (blue circles) and they vanish above this threshold (green triangles).}
	\label{fig:cr_sf}
\end{figure}

Without exponential cutoff, the probability distribution and double generating function of a scale-free network \cite{goh01,chung02} are the following:
\eq{p_{ij}=\frac{(ij)^{-\gamma}}{\zeta(\gamma)^2}}{eq:SF_pij}
\eq{g_{00}(x,y)=\frac{\mathrm{Li}_\gamma(x)\mathrm{Li}_\gamma(y)}{\zeta(\gamma)^2}}{eq:SF_g00}
and the generating function of the excess degree distributions:
\eq{g_{10}(x,y)=\frac{\mathrm{Li}_{\gamma-1}(x)\mathrm{Li}_\gamma(y)}{\kav x\zeta(\gamma)^2}}{eq:SF_g10}
\eq{g_{01}(x,y)=\frac{\mathrm{Li}_\gamma(x)\mathrm{Li}_{\gamma-1}(y)}{\kav y\zeta(\gamma)^2}}{eq:SF_g01}
Newman showed in \cite{newman10} that the condition for the existence of the giant components is
\eq{\partial_x\partial_yg_{00}(x,y)|_{x,y=1}>\kav}{eq:SF_giant_comp_condition1}
For the scale-free network, this equation reads as
\eq{\frac{\mathrm{Li}_{\gamma-1}(x)\mathrm{Li}_{\gamma-1}(y)}{xy\zeta(\gamma)^2}\bigg|_{x,y=1}>\kav}{eq:SF_giant_comp_condition2}
Now, if make use of the relation between the exponent and the average degree: $\kav=\frac{\zeta(\gamma-1)}{\zeta(\gamma)}$, and substitute $x=1$, $y=1$ we get:
\eq{\kav^2>\kav}{eq:SF_giant_comp_condition3}
or equivalently
\eq{\kav>1}{eq:SF_giant_comp_condition4}
So, there is a giant component if the average degree is larger than one. But if we look at the function
\eq{\frac{\zeta(\gamma-1)}{\zeta(\gamma)}}{eq:SF_k_av}
we can conclude that this condition gives giant components for any $\gamma\ge2$. Numerical simulations show that this is not the case (see Fig.~\ref{fig:cr_sf}).

If we substitute $g_{00}(x,y)$ and its derivatives in Eq.~(\ref{eq:grc_def2}) and Eq.~(\ref{eq:ER_u_eq})-(\ref{eq:ER_v_eq}) we observe that there are giant components of unit size for every exponent larger than two, since the formula
\eqarray{\pi_s^{out}	& = & \frac{\kav}{(s-1)!}\Bigg[\frac{\dif^{s-2}}{\dif y^{s-2}}\bigg(g_{01}(1,y)g_{10}(1,y)^{s-1}\bigg)\Bigg]_{y=0} \nonumber \\
			& = & \frac{1}{\zeta(\gamma)^s(s-1)!}\Bigg[\frac{\dif^{s-1}}{\dif y^{s-1}}\bigg(\mathrm{Li}_\gamma(y)^s\bigg)\Bigg]_{y=0}}{eq:SF_pi}
gives exactly zero for any $s$ and the equation
\eq{u=g_{01}(u,1)=\frac{\mathrm{Li}_\gamma(u)}{\zeta(\gamma)}}{eq:SF_u_eq}
has always two solutions: $u_1=0$ and $u_2=1$.

These show the limitations of the generating function formalism in directed networks. The limits of the method is already well-known in some cases of undirected scale-free networks as well \cite{newman03}. In the undirected case, networks with small exponents tend to have many hubs, and the clustering coefficient also increases remarkably \cite{newman03}. Large clustering coefficient means, that the graph is not locally tree-like, which is the main assumption of the applied method. Eq.~(\ref{eq:SF_pi})-(\ref{eq:SF_u_eq}) point out that, for directed networks, the method can be barely applied to scale-free networks.

\begin{figure}
    \centering
	\includegraphics[angle=-90,width=10cm]{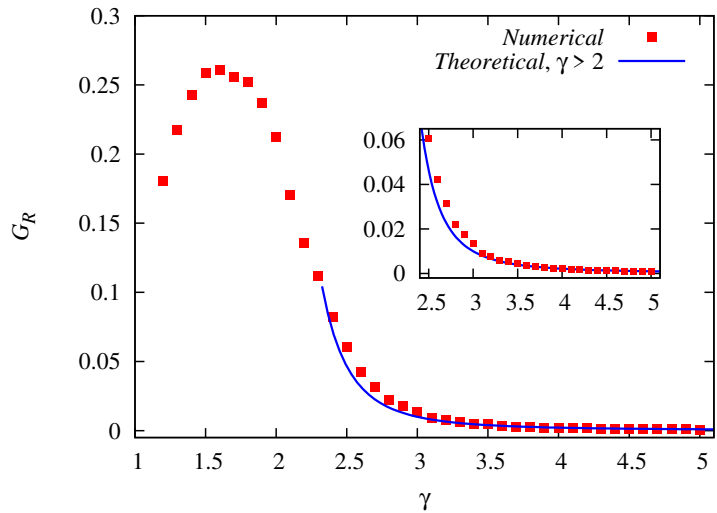}
	\caption{Global reaching centrality of the scale-free graph. Numerical results are the average of 1000 independent calculations on networks with $N=10^4$. Errors have a magnitude of the dot sizes. The largest degree approximation fits well above $\gamma>3$ and gives a lower value below it.}
	\label{fig:grc_sf}
\end{figure}

However, it is possible to give a qualitative approximation for the $G_R$ in the case of $\gamma>2$ (this is the most important regime in terms of real networks \cite{barabasi02}). We use our observation from the simulations that there is no $\mathcal{G}_{S}$, and that in scale-free networks, very large degrees can appear. Since a large amount of the nodes have few out-degrees, the $G_R$ is obviously dominated by $C_R^{max}$. Let us assume that the network breaks down into small components whose are the neighborhoods of the nodes with large out-degrees (i.e., every component gathers around a hub). In this case, the largest reachable set is the largest out-degree. In a scale-free network with degree distribution of $p_k\propto k^{-\gamma}$, the largest degree is well approximated by $k_{max}\approx N^{\frac{1}{\gamma-1}}$ \cite{cohen00}. Using this approximation for the out-degrees and not taking into account the in-degrees we get:
\eq{G_R\simeq N^{\frac{2-\gamma}{\gamma-1}}}{eq:SF_grc}
Comparison of the real $G_R$ and this approximation is shown in Fig.~\ref{fig:grc_sf}.
For large exponents, Eq.~(\ref{eq:SF_grc}) fits well to the numerical results. Below $\gamma=3$, it becomes less accurate. Note that the lower the exponent is the more portion of the nodes has large degrees. The predicted $G_R$ becomes larger than the real value at some point ($\gamma\approx2.3)$, that is the number of large hubs increases which results in larger average reaching centrality as well.

It is worth to stop here and understand the non-monotonic dependence of the $G_R$ on the exponent. For large exponents, the scale-free network tends to behave similar as the Erd\H{o}s--R\'{e}nyi graph. Thus, it becomes more homogeneous. Its average degree also decreases and instead of the hubs, more and more small, disconnected components appear. As the network becomes more fragmented, both the average and maximum reaching centrality decreases, which results in a very small $G_R$. In the case of small exponents, it is known that the clustering coefficient of undirected scale-free networks significantly increases \cite{newman03}. This phenomenon is likely to be present in the directed case as well. With large clustering, the number of \emph{directed circles} also increases. However, the nodes in a directed circle have exactly the same reachable set, and thus similar local reaching centrality. This tendency of equalization in the local reaching centralities affects the heterogeneity of the distribution, and enables multiple nodes to reach (or approach closely) the maximum reaching centrality. The result is again a decreasing $G_R$.

\section{Effects of correlations}
\subsection{Degree-correlations}
In the above sections we assumed that the joint probability distribution of the in- and out-degrees is simply the product of two independent distributions, i.e., there are no degree-correlations. However, it is known that there are such correlations in real networks \cite{newman03b}. It is also known that these correlations have markable effects on the different properties of networks (percolation thresholds, epidemic thresholds, etc) \cite{newman02,boguna03}. In this paper, we numerically study the effect of two types of correlations on the $G_R$: one-point correlations and directed assortative mixing. One-point correlation is the Pearson correlation between the in- and out-degree of the nodes:
\eq{\rho=\frac{\langle k_{in}k_{out}\rangle_V-\langle k_{in}\rangle_V\langle k_{out}\rangle_V}{\sigma_V(k_{in})\sigma_V(k_{out})}}{eq:Corr_rho_def}
The brackets denote averages over the nodes and $\sigma_V(k)=\sqrt{\langle k^2\rangle_V-\langle k\rangle_V^2}$ is the standard deviation of the corresponding degrees. Somewhat similarly, one can define the Pearson correlation between the out-degree of the start of an edge and the in-degree of its end term \cite{newman03b}:
\eq{r=\frac{\langle j_{in}k_{out}\rangle_E-\langle j_{in}\rangle_E\langle k_{out}\rangle_E}{\sigma_E(j_{in})\sigma_E(k_{out})}}{eq:Corr_r_def}
Here $j_{in}$ is the in-degree of the node an edge points to and $k_{out}$ is the out-degree of the node that edge comes from. The averages run over the edges. We will refer to this quantity as two-point correlation or directed assortativity, since it is a plausible generalization of the assortativity in undirected networks. Table~\ref{tab:real_network_correlations} shows the two correlations in several real networks.
\begin{table}
    \centering
    \begin{tabular}{lccc}
	    Network & $\kav$ & $\rho$ & $r$ \\
	    \hline
	    \multicolumn{3}{l}{\bf Food networks} & \\
	    Ythan \cite{dunne02}			& 4.452 & 0.168 & -0.249 \\
	    LittleRock \cite{martinez91}		& 13.628 & -0.138 & -0.394 \\
	    Grassland \cite{dunne02}		& 1.557 & -0.179 & -0.233 \\
	    \multicolumn{3}{l}{\bf Electric} & \\
	    s1488 \cite{circuits}			& 2.085 & -0.274 & 0.218 \\
	    s5378 \cite{circuits}			& 1.467 & -0.137 & 0.151 \\
	    s35932 \cite{circuits}			& 1.683 & -0.074 & 0.088 \\
	    \multicolumn{3}{l}{\bf Trust} & \\
	    WikiVote \cite{leskovec10}		& 14.573 & 0.318 & -0.083 \\
	    College	\cite{vanduijn03,milo04}	& 3 & 0.053 & -0.159 \\
	    Prison	\cite{vanduijn03,milo04}	& 2.716 & 0.201 & 0.129 \\
	    \multicolumn{3}{l}{\bf Regulatory} & \\
	    TRN-Yeast-1 \cite{balaji06}		& 2.899 & 0.025 & -0.173 \\
	    TRN-Yeast-2 \cite{milo02}		& 1.568 & -0.236 & -0.220 \\
	    TRN-EC \cite{milo02}			& 1.239 & -0.082 & 0.085 \\
	    \multicolumn{3}{l}{\bf Metabolic} & \\
	    \emph{C. elegans} \cite{jeong00}	& 2.442 & 0.924 & -0.174 \\
	    \emph{E. coli} \cite{jeong00}		& 2.533 & 0.923 & -0.167 \\
	    \emph{S. Cerevisiae} \cite{jeong00}	& 2.537 & 0.923 & 0.182 \\
    \end{tabular}
    \caption{The one- and two-point correlations of real networks along with their references. With the exception of the metabolic networks, most of them have small correlations.}
    \label{tab:real_network_correlations}
\end{table}

\subsection{One-point correlations}
\subsubsection{Numerical results}
\begin{figure}
    \centering
	\mbox{
	\subfloat[\textbf{Erd\H{o}s--R\'{e}nyi}]{\includegraphics[angle=-90,width=8cm]{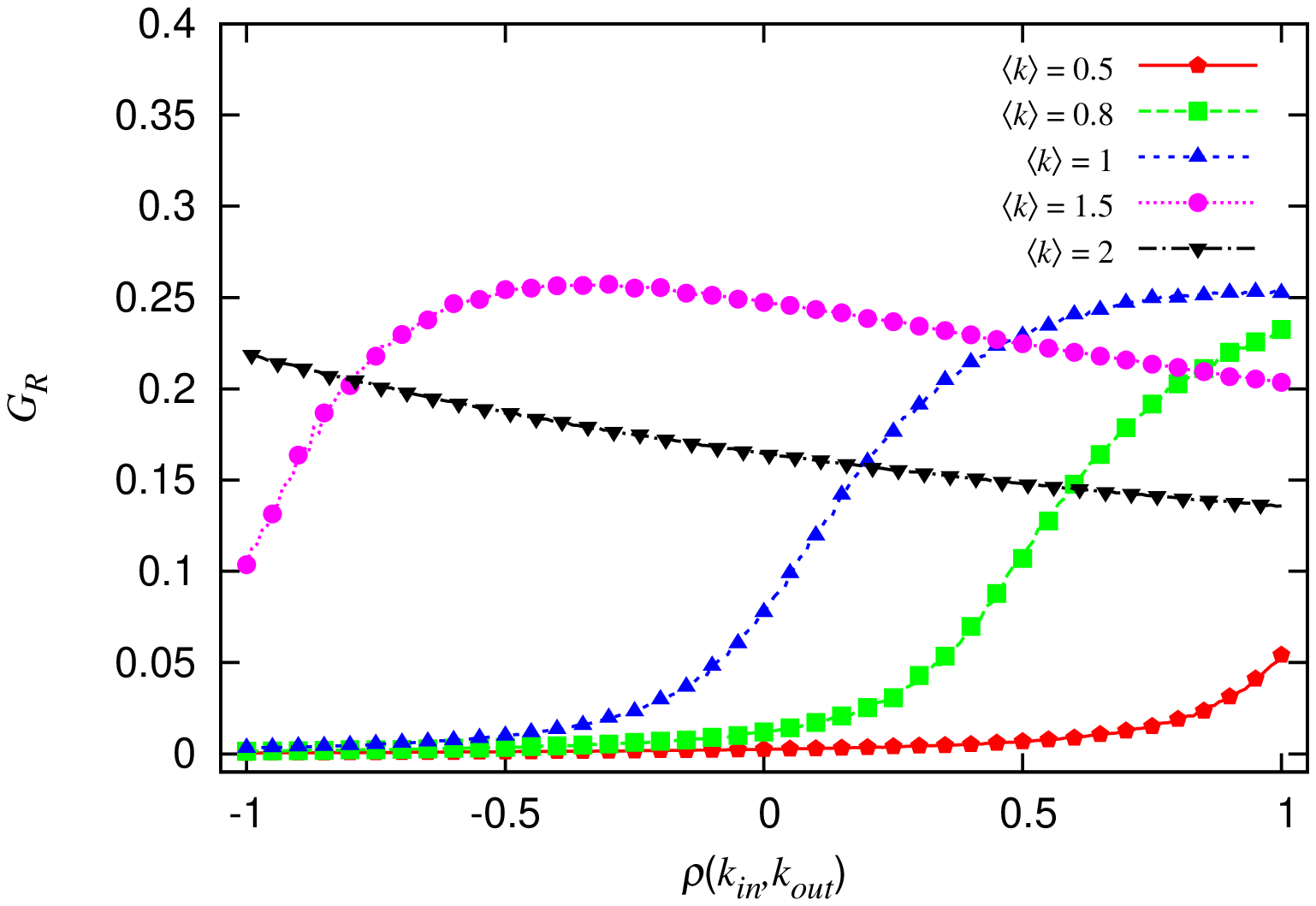}}
	\subfloat[\textbf{Exponential}]{\includegraphics[angle=-90,width=8cm]{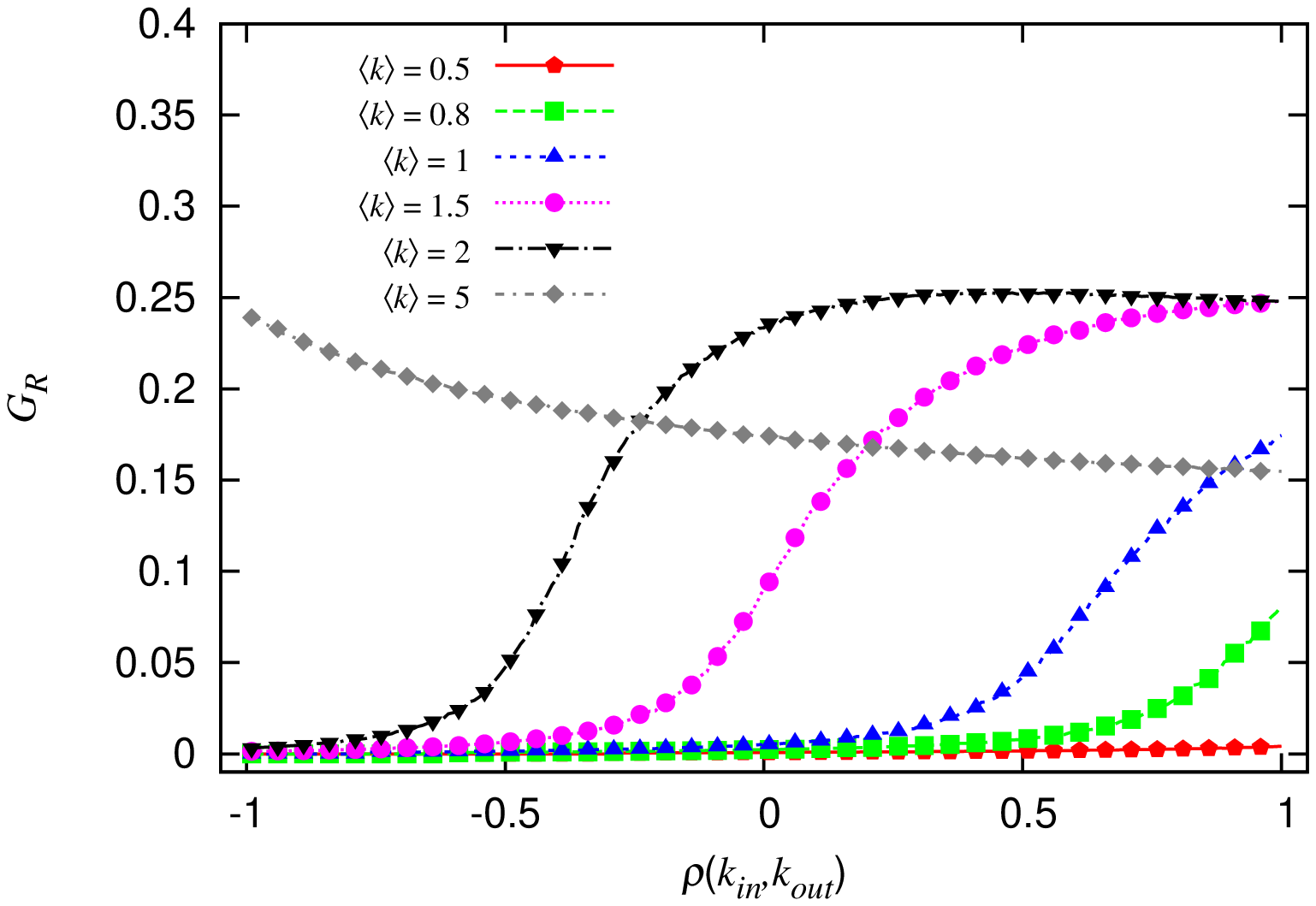}}}
	\mbox{
	\subfloat[\textbf{Scale-free}]{\includegraphics[angle=-90,width=8cm]{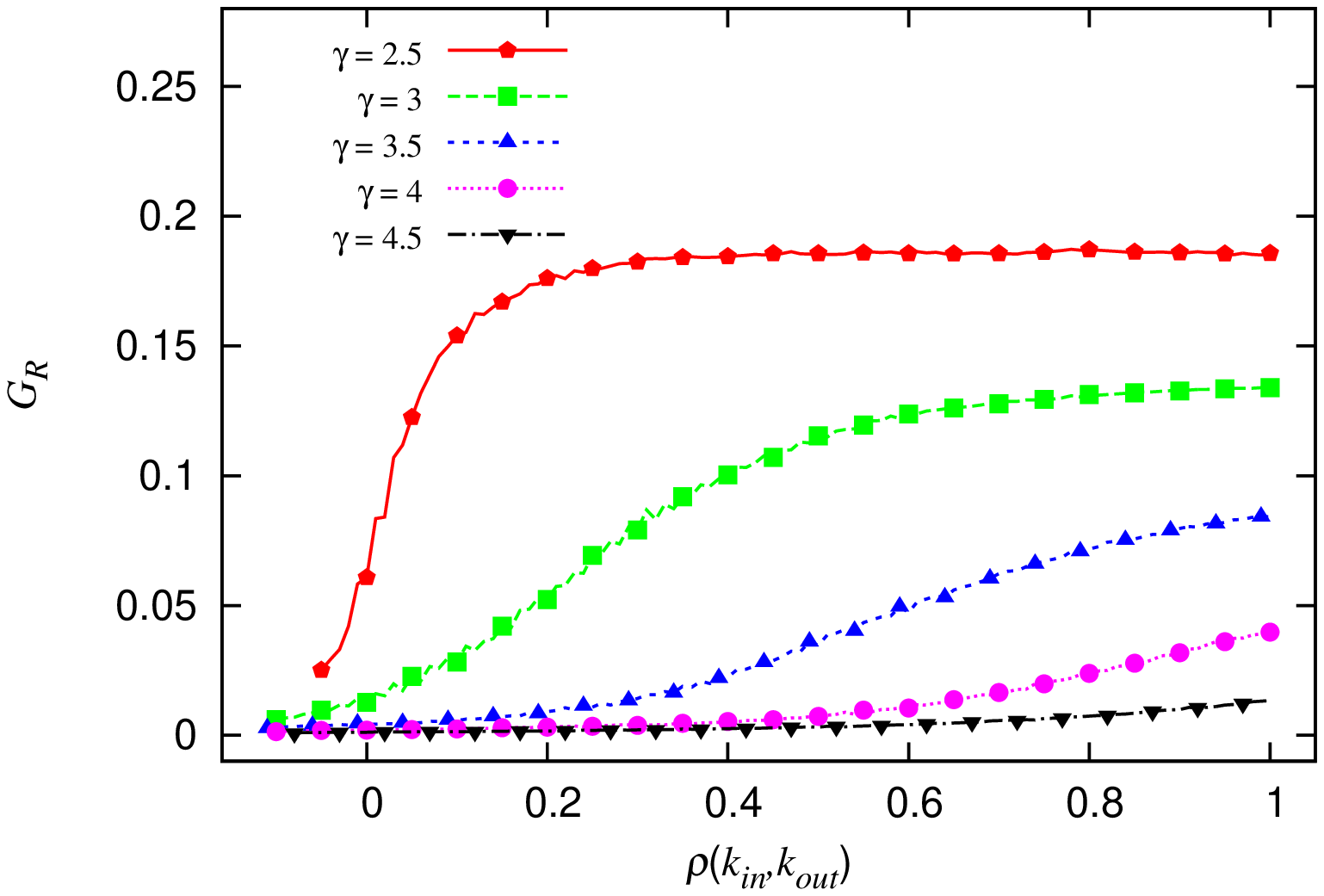}}}
	\caption{The $G_R$ versus the one-point correlations for different average degrees in Erd\H{o}s--R\'{e}nyi and exponential graph and for different exponents in the scale-free network. Every point on the plots is the average of at least 200 independent simulations on networks with $N=10^4$ nodes. In the case of the scale-free network, only a small negative degree-correlation is accessible through random optimization.}
	\label{fig:one_point_corr}
\end{figure}

In order to study the effect of the one-point correlation, we generated the in- and out-degree lists for the network with a given distribution and average degree. After randomizing both lists, we fixed the out-degree list while we successively swapped randomly chosen elements in the in-degrees. After every swap, we calculated $\rho(k_{in},k_{out})$ and accepted the new list whenever the correlation increased (decreased). We measured the $G_R$ when the difference between the correlation of the current state and the last measured state was larger than 0.01 (this is the resolution of the measured $G_R(\rho)$ function). The dependence of the $G_R$ on the one-point correlation is shown in Fig.~\ref{fig:one_point_corr}.
It is clearly seen that the behavior of the $G_R$ varies and depends strongly on the average degree. In the ER and exponential graphs, even the monotonity of the curves change. In the scale-free network, the $G_R$ saturates at some level of correlation and the exponent effects only the threshold above which further correlations do not change the $G_R$. 

\subsubsection{A qualitative approach}
In this section, we present a simple argument to understand the direction of change if small correlations are present in the graph. It is important to emphasize, that the analytic expression for the effects of correlations is out of the scope of this paper and we intend to give only a rather qualitative insight of the behavior. For this, let us consider the following joint degree distribution:
\eq{p_{jk}=p_jp_k+\rho\sigma_p^2m_{jk}}{eq:1pCorr_modified_pij}
Where the only criteria for the $m_{jk}$ matrix are:
\eq{\sum_{j=0}^\infty m_{jk}=\sum_{k=0}^\infty m_{jk}=0}{eq:1pCorr_mij_crit1}
\eq{\sum_{j,k=0}^\infty jkm_{jk}=1}{eq:1pCorr_mij_crit2}
With these conditions, the one-point degree-correlation of the above joint degree distribution is exactly $\rho$ \cite{newman03b}. Note that we assumed the same distribution for the in- and out-degrees ($\sigma_V(k_{in})=\sigma_V(k_{out})=\sigma_p$). The generating function of the modified distribution:
\eq{g_{00}(x,y)=g_{00}^p(x,y)+\rho\chi_{00}(x,y)}{eq:1pCorr_modified_g00}
where the $g_{00}^p$ is the generating function of the original joint degree distribution and we introduced the following function:
\eq{\chi_{00}(x,y)=\sigma_p^2\sum_{j,k=0}^\infty x^jy^km_{jk}}{eq:1pCorr_khi00_def}
Above the critical average degree, we have to solve
\eq{u=g_{01}^p(u,1)+\frac{\rho}{\kav}\partial_y\chi_{00}(u,1)}{eq:1pCorr_modified_u_eq}
and the $G_R$ is given by the modified generating function:
\eq{G_R=g_{00}(u,1)-g_{00}(u,u)}{eq:1pCorr_modified_grc}
The exact solution of Eq.~(\ref{eq:1pCorr_modified_u_eq}) depends on the actual choice of $m_{jk}$ and in most cases not possible to find in closed form. However, for small correlations ($\rho\ll1$), we can describe the change in the $G_R$ at given average degree. In this case, the solution of Eq.~(\ref{eq:1pCorr_modified_u_eq}) is very close to the solution of the original equation without the correlations: $u=u_0+u_\rho$. Expanding both sides and keeping only the linear terms in $\rho$ and $u_\rho$, we get:
\eq{u_\rho=\frac{1}{\kav}\frac{\partial_y\chi_{00}(u_0,1)}{1-\partial_xg_{01}^p(u_0,1)}\rho=\beta(u_0)\rho}{eq:1pCorr_urho_def}
The coefficient $\beta(u)$ is negative above the transition point for the model networks we are interested in (for the proof, see Appendix B). In the linear approximation, the $G_R$ looks like:
\eq{G_R=G_R^{(0)}+G_R^{(\rho)}}{eq:1pCorr_grc_result}
where
\eq{G_R^{(0)}=g_{00}^p(u_0,1)-g_{00}^p(u_0,u_0)}{eq:1pCorr_grc0_def}
and
\eq{G_R^{(\rho)}=\bigg[\beta(u_0)(1-2u_0)\cdot\frac{\dif g_0^p(x)}{\dif x}\Big|_{x=u_0}-\chi_{00}(u_0,u_0)\bigg]\rho}{eq:1pCorr_grcrho_def}
In the derivation of $G_R^{(\rho)}$, we used the observations that when the joint degree distribution factorizes, the generating function is also a product of the single distributions, i.e.,
\eq{g_{00}^p(x,y)=g_0^p(x)g_0^p(y)}{eq:1pCorr_g00p_factoring}
and also
\eq{g_0^p(u_0)=\sum_{j=0}^\infty p_ju_0^j=\frac{1}{\kav}\sum_{j,k=0}^\infty ku_0^jp_{jk}=g_{01}^p(u_0,1)=u_0}{eq:1pCorr_g0p_equal_u0}
Now we have to interpret the result we obtained for the change in the $G_R$. First, for simplicity, let us choose the $m_{jk}$ as proposed in \cite{newman03b}:
\eq{m_{jk}=\frac{(p_j-\phi_j)(p_k-\phi_k)}{\big(\kav-\kav_\phi\big)^2}}{eq:1pCorr_mij_example}
where $\phi_j$ is an arbitrary normalized distribution and $\kav_\phi$ is its average. Moreover, choose the $\phi_j$ distribution such that $\kav_\phi=1$ which is the critical point of the ER and exponential networks. In this case $\chi_{00}(u_0,u_0)\ge0$ and also note that
\eq{\beta(u_0)<0}{eq:1pCorr_beta_negative}
\eq{\frac{\dif g_0^p(x)}{\dif x}\Big|_{x=u_0}>0}{eq:1pCorr_dg0p_dx_positive}
above the transition point. The behavior of $G_R^{(\rho)}$ is governed by the relation of the two terms in the right hand side of Eq.~(\ref{eq:1pCorr_grcrho_def}). It is easy to check that near the critical point ($u_0\approx1$), the second term is very close to zero, and $G_R^{(\rho)}$ is dominated by the first term which is positive. This means that when the average degree is small, but the graph already has giant components of finite size, small one-point correlations increase its hierarchical structure, as in Fig.~\ref{fig:one_point_corr}. For large average degrees, $u_0\to0$ and both terms in Eq.~(\ref{eq:1pCorr_grcrho_def}) becomes negative, resulting in a decreasing effect of small correlations, in good accordance with the numerical results (see Fig.~\ref{fig:one_point_corr}a and Fig.~\ref{fig:one_point_corr}b). In the regime where the $G_R$ has a maximum ($u_0=\frac{1}{2}$), Eq.~(\ref{eq:1pCorr_grcrho_def}) predicts a negative coefficient for $\rho$, but as Fig.~\ref{fig:one_point_corr}b shows, this is not the case for the exponential network. This discrepancy suggests that the change in the slope near this point is not trivial and strongly depends on the details of the addition of correlations.

Below the transition point, there are no giant components and the $G_R$ is dominated by the largest reachable set.
\begin{figure*}
    \centering
	\subfloat[]{\includegraphics[width=5cm]{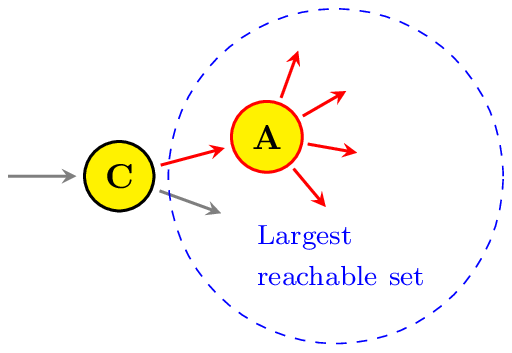}}
	\subfloat[]{\includegraphics[width=5cm]{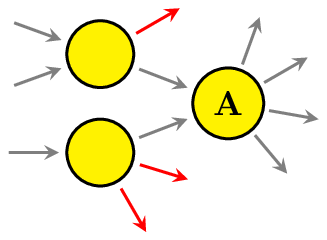}}
	\subfloat[]{\includegraphics[width=5cm]{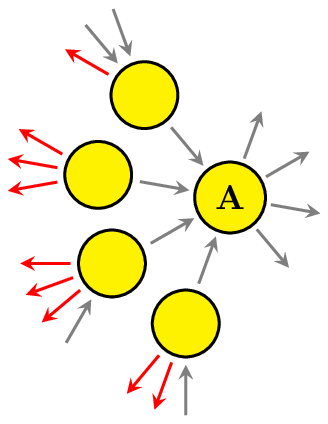}}
	\caption{The largest reachable set below the transition point (a) and one of the nodes with large out-degree in the largest reachable set (\textbf{A}). Without degree-correlations, the expected number of in-degrees of \textbf{A} is the average degree and the maximum out-degree of its in-neighbors is also moderate (b). With degree-correlations, the nodes with large out-degree has large in-degree as well (c). And with increasing number of in-neighbors the expected maximum out-degree of its neighbors increases as well.}
	\label{fig:one_point_corr_below_transition}
\end{figure*}
We can assume that the node with the largest reachable set (node \textbf{C} on Fig.~\ref{fig:one_point_corr_below_transition}a) or some of its out-neighbors have large out-degree, since this increases the probability of reaching more other nodes. Consider this neighbor with the large out-degree (node \textbf{A} in Fig.~\ref{fig:one_point_corr_below_transition}a). In the case of negative or zero correlations, the expected maximum of the out-degree of its in-neighbors is small (Fig.~\ref{fig:one_point_corr_below_transition}b). But in the presence of positive one-point correlations, the number of in- and out-edges tend to be similar, in other words, the node in question has large in-degree as well. In this case, the expected maximum of the out-degree of its in-neighbors (and thus of the candidate node for the largest local reaching centrality) increases and there are more reachable nodes. The scale-free network is a special case in the sense that the largest reachable set is with good approximation the neighborhood of a hub. Because of the correlations, this hub tends to have many in-neighbors as well and it is more likely to have a neighbor with many links (Fig.~\ref{fig:one_point_corr_below_transition}d). The numerical results in Fig.~\ref{fig:one_point_corr}c suggest that the maximum out-degree of the nearest neighbors reaches its saturation level at relatively small correlations.

\subsection{Two-point correlations}
\begin{figure}
    \centering
	\mbox{
	\subfloat[\textbf{Erd\H{o}s--R\'{e}nyi}]{\includegraphics[angle=-90,width=8cm]{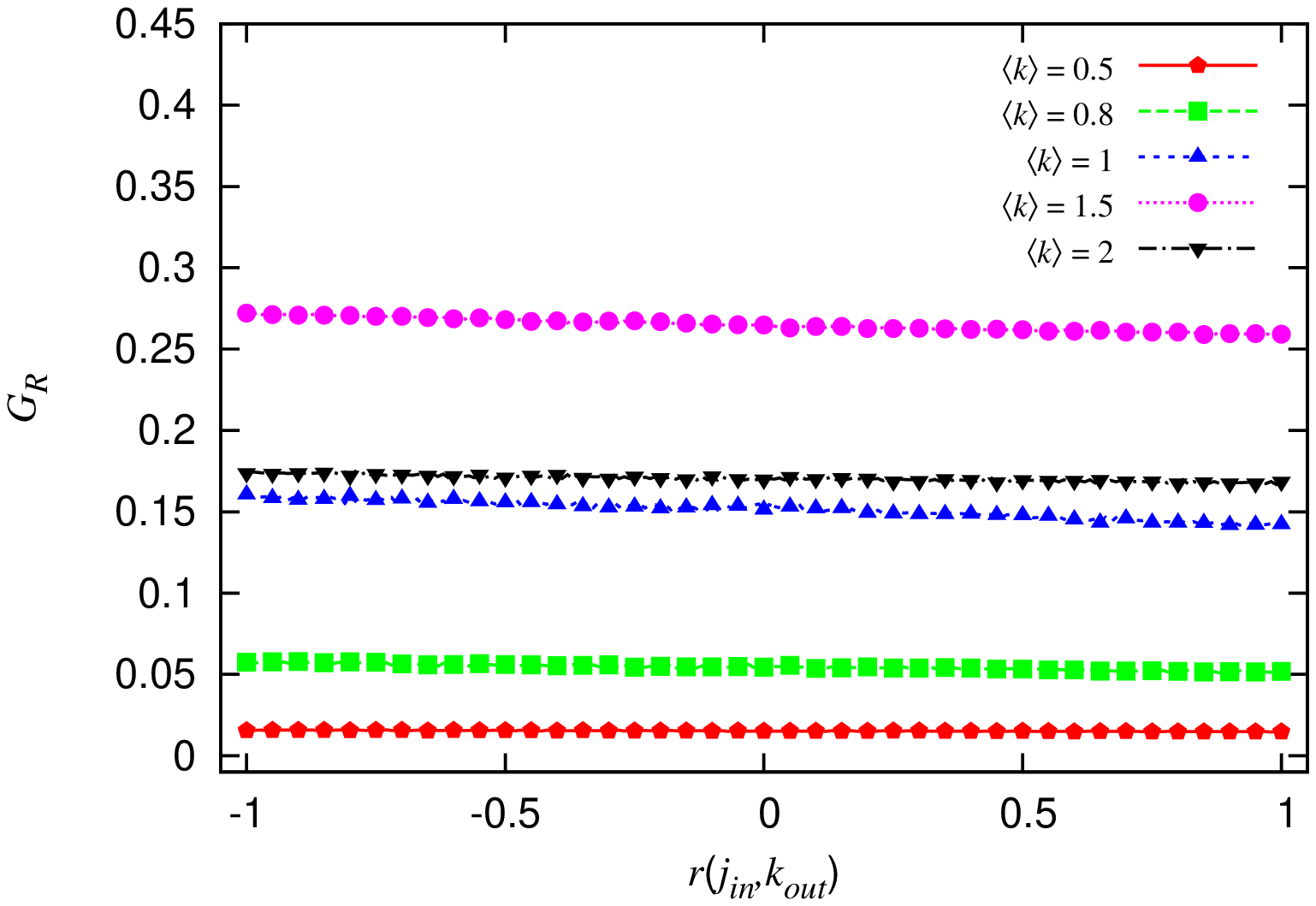}}
	\subfloat[\textbf{Exponential}]{\includegraphics[angle=-90,width=8cm]{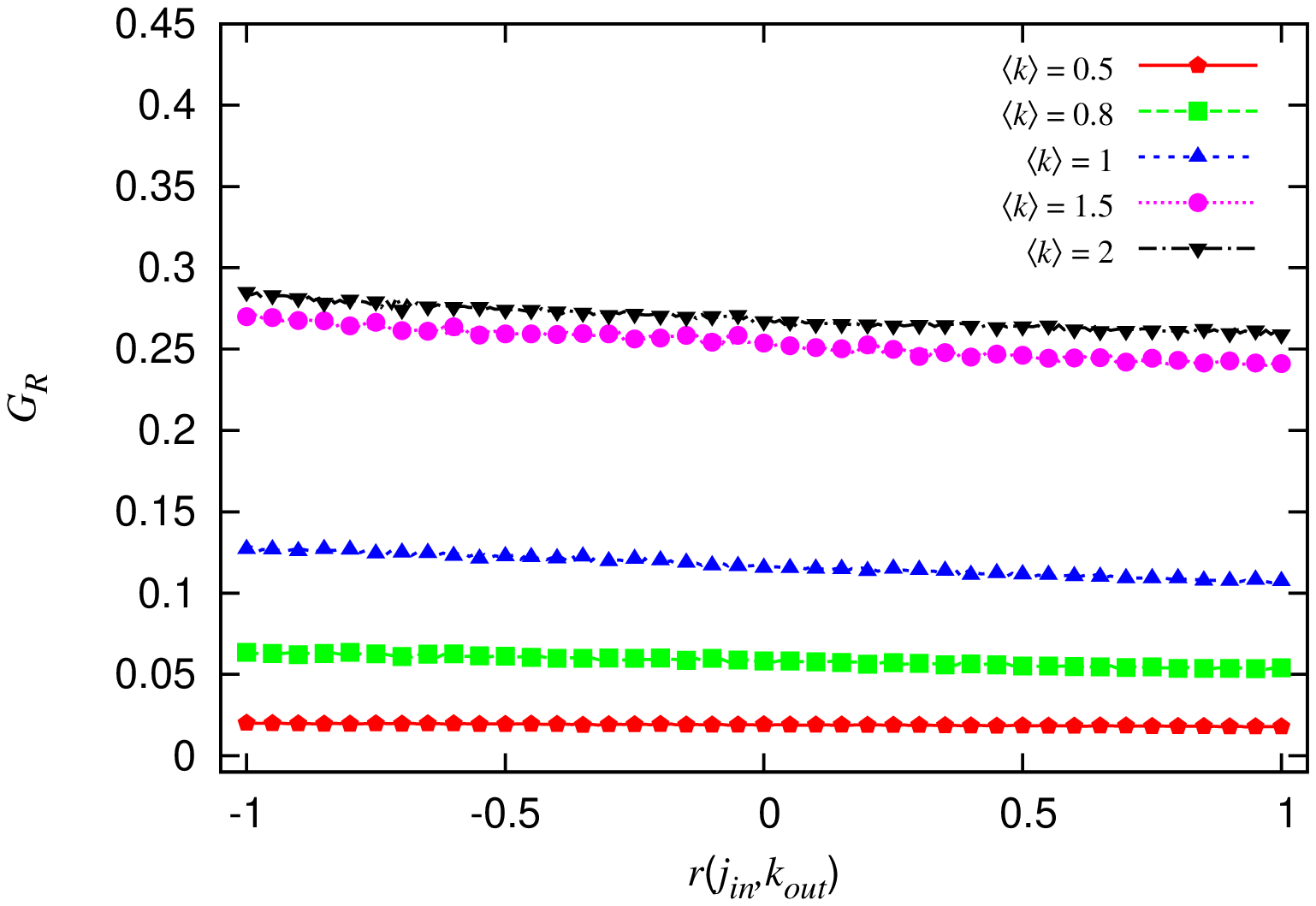}}}
	\mbox{
	\subfloat[\textbf{Scale-free}]{\includegraphics[angle=-90,width=8cm]{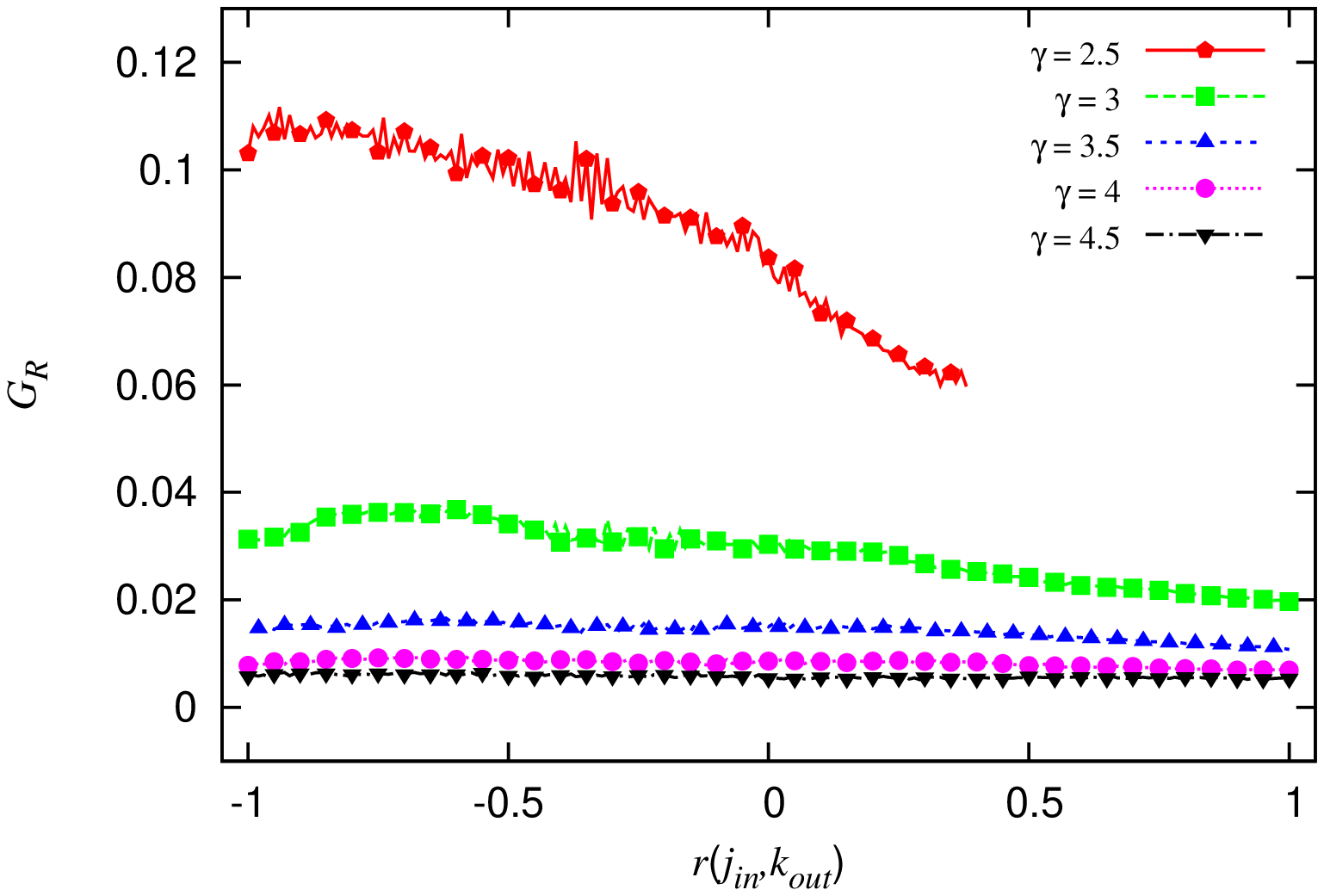}}}
	\caption{The effect of two-point correlations to the $G_R$ for the random graph models. Every point is an average at least 100 independent measure on networks with $N=1000$.}
	\label{fig:two_point_corr}
\end{figure}

\begin{figure}
    \centering
	\mbox{
	\subfloat[$\boldsymbol{r}\boldsymbol{>}\boldsymbol{0}$]{\includegraphics[width=6cm]{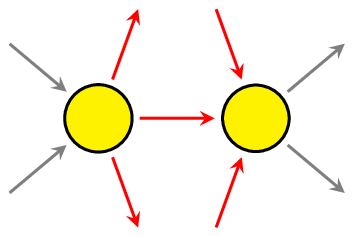}}
	\subfloat[$\boldsymbol{r}\boldsymbol{>}\boldsymbol{0}$]{\includegraphics[width=6cm]{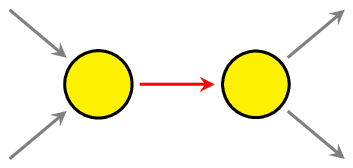}}}
	\mbox{
	\subfloat[$\boldsymbol{r}\boldsymbol{<}\boldsymbol{0}$]{\includegraphics[width=6cm]{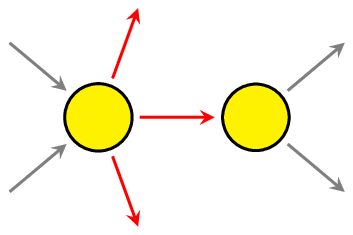}}
	\subfloat[$\boldsymbol{r}\boldsymbol{<}\boldsymbol{0}$]{\includegraphics[width=6cm]{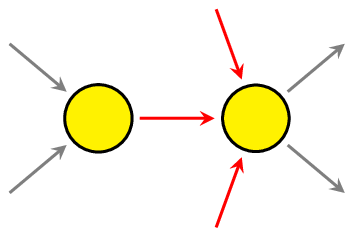}}}
	\caption{The illustration of the two limiting cases of two-point correlations $r$ defined by Eq.~(\ref{eq:Corr_r_def}). The edges that are correlated are shown in red. In the case of $r>0$, the reachable sets of hubs tend to be similar (a) and bottlenecks emerge (b), causing an increase in the overlap of reachable sets. In negatively correlated networks ($r<0$), hubs can have their own reachable sets (c) and nodes with large in-degree tend to be reachable from independent parts of the network (d), decreasing the overlap between reachable sets.}
	\label{fig:two_point_corr_explain}
\end{figure}

The two-point correlation of the type we investigate in this paper is the generalization of the assortativity to directed networks \cite{newman03b}. In order to introduce this correlation in our calculations, let $\mathcal{P}(j_{in},j_{out};k_{in},k_{out})$ denote the probability that a randomly chosen edge starts at a node with $j_{in}$ in- and $j_{out}$ out-degree and points to a node with $k_{in}$ in- and $k_{out}$ out-degree. This probability satisfies the following equations:
\begin{equation}
    \sum_{j_{in},j_{out}=0}^\infty\mathcal{P}(j_{in},j_{out};l_{in},l_{out})=\sum_{k_{in},k_{out}=0}^\infty\mathcal{P}(l_{in},l_{out};k_{in},k_{out})=p_{l_{in}l_{out}}
    \label{eq:2pCorr_P_to_pij}
\end{equation}
which also create its connection to the joint degree distribution. Without two-point correlations (and other correlations), the two-node joint distribution factorizes:
\eq{\mathcal{P}^0(j_{in},j_{out};l_{in},l_{out})=p_{j_{in}}p_{j_{out}}p_{k_{in}}p_{k_{out}}}{eq:2pCorr_P_without_corr}
We can modify it by adding a correlation between $p_{j_{out}}$ and $p_{k_{in}}$:
\begin{equation}
    \mathcal{P}^r(j_{in},j_{out};k_{in},k_{out})=p_{j_{in}}p_{k_{out}}\big[p_{j_{out}}p_{k_{in}}+r\sigma(p_{j_{out}})\sigma(p_{k_{in}})m_{j_{out}k_{in}}\big]
    \label{eq:2pCorr_P_with_corr}
\end{equation}
However, if we recover the joint degree distribution, the added term vanishes:
\eqarray{p^r_{l_{in}l_{out}}	& = & \sum_{j_{in},j_{out}=0}^\infty\mathcal{P}_r(j_{in},j_{out};l_{in},l_{out}) \nonumber \\
				& = & p^0_{l_{in}l_{out}}}{eq:2pCorr_pijr_equals_pij0}
because the sum of any row or column of the matrix $m_{j_{out}k_{in}}$ must be zero (Eq.~(\ref{eq:1pCorr_grcrho_def})). This means that the assortativity defined by Eq.~(\ref{eq:Corr_r_def}) does not affect the $G_R$ directly (via the joint degree distribution). Fig.~\ref{fig:two_point_corr} depicts the numerical results that were produced by the same protocol as in the previous subsection.
In the case of the ER and exponential networks, a small decrease in the $G_R$ can be observed, regardless of the average degree. The SF network has a non-monotonous behavior, but for larger assortativity, a decreasing in the $G_R$ can be seen. These results point out that, although two-point correlation does not affect the joint degree distribution directly, it changes the hierarchical structure of the network a little. This small effect can be understood if we look at the different typical structures around an edge (Fig.~\ref{fig:two_point_corr_explain}).
In the case of large two-point correlations, the nodes with large out-degree (those that are expected to have larger reachable set) have out-neighbors with also large in-degrees (Fig.~\ref{fig:two_point_corr_explain}a). This means that the nodes they can directly reach are also reachable from many other nodes (that also have large out-degrees). This reduces the difference between the local reaching centralities among nodes with large $C_R$ or, in other words, introduces higher $c_R(i)$ in the definition of $G_R$ (see Eq.~(\ref{eq:grc_def1})). Positive two-point correlation also introduces bottlenecks in the network (Fig.~\ref{fig:two_point_corr_explain}b). These structures accumulate many nodes on the in-edges that has similar reachable set, thus also reducing the differences. On the other hand, if the graph has negative two-point correlations, nodes with large out-degree can reach their out-neighbors uniquely (Fig.~\ref{fig:two_point_corr_explain}c). Nodes that are reachable from independent directions also emerge (Fig.~\ref{fig:two_point_corr_explain}d). These nodes have in-neighbors with small out-degree. Both effects decrease the ratio of overlapping reachable sets which results in a reduced similarity in the $C_R$ values.

\section{Discussion}
The global reaching centrality ($G_R$) is a measure constructed to provide a characteristic number for the flow hierarchy of a directed network. In this paper we investigated the behavior of this measure and its dependence on the joint degree distribution. We have shown that the hierarchical structure of a random network strongly depends on the existence of the giant strongly connected component ($\mathcal{G}_{S}$) and other giant components in the network. In the regime of low average degree, the $G_R$ is dominated by the node with the largest reachable set that can be approximated from the distribution of the local reaching centralities. This distribution is connected to the distribution of small out-components in a network and can be analytically determined using the generating function method generalized to directed networks. In the presence of the $\mathcal{G}_{S}$, the $G_R$ can be well approximated by the difference in the sizes of the giant in-component ($\mathcal{G}_{in}$) and the $\mathcal{G}_{S}$, and both can be calculated exactly for some network models with given degree distribution. Using the approximations for the $G_R$, we calculated its dependence on the average degree for different network models: a directed random tree, the Erd\H{o}s--R\'{e}nyi graph (ER), the exponential the scale-free (SF) network. The results show that in the sense of hierarchy, there is an optimal average degree at which the $G_R$ has a maximum value. It is a result of the competition between the size of the largest reachable set (which can be found by nodes in the $\mathcal{G}_{in}$) and the average size of reachable sets (which is dominated by the nodes in the $\mathcal{G}_{S}$). Near to the transition point, the size of the $\mathcal{G}_{in}$ is close to the size of the $\mathcal{G}_{S}$ and the difference is increasing since the $\mathcal{G}_{in}$ grows faster in the beginning. In the limit of large average degree, the sizes of both giant components tend to become equal, since even the two set of nodes tend to be the same.

We also investigated the dependence of the $G_R$ on two types of degree-correlations: one-point correlation (in-degree and out-degree of a node) and two-point correlation (the out-degree of the source and the in-degree of the target of a directed edge). Numerical results show that the hierarchical structure changes systematically with the one-point correlation. This is in accordance with the fact that the $G_R$ can be expressed with the joint degree distribution, and latter is directly affected by the one-point correlations. We have pointed out that in the two limiting case of the average degree, the qualitative effect of small correlations can be understand by the addition of a correlation term to the uncorrelated joint degree distribution. When the average degree is very low, small correlations have a positive effect on the $G_R$. This is not true for denser graph, in which correlations decrease the $G_R$. There is also a regime of the average degree in which a given level of correlations can maximize the hierarchy.

Both numerical and analytical results suggest that two-point correlation (which is the generalization of the assortativity for directed networks) does not effect the $G_R$ directly. However, a small negative effect can be observed. It can be understood by looking at the effect of the two-point correlation on the neighborhood of an edge.

The results on the random network models have shown a deeper insight of the behavior of the $G_R$ but we have to keep in mind that they are only the first steps in the understanding of hierarchy. The main message of the results is that hierarchy is sensible to the edge density of a network and tends to emerge more likely in sparse networks. This is in good accordance with the observation that most of the real world networks are in the range of low density and many of them has an inherent hierarchical structure. The results also point out that correlations can have a large effect and can change the the hierarchical structures fundamentally (they can produce a finite magnitude of $G_R$ even if the network would have an infinitesimal $G_R$ otherwise). This is another indicator of the well-known fact that one has to take into account the presence of correlations when dealing with real networks.

\section*{Acknowledgement}
The author would like to thank Tam\'{a}s Vicsek, P\'{e}ter Pollner and Gergely Palla for the fruitful conversations that facilitated the interpretation of the results and for their advices during the preparation of the manuscript. This research was fully supported by the EU ERC FP7 COLLMOT Grant No: 227878.

\appendix
\section{Out-components in directed graphs}
Using the definition of the generating function $h_0(y)$, we can obtain the probabilities for the out-components:
\eq{\pi_s^{out}=\frac{1}{(s-1)!}\Bigg[\frac{\dif^{s-1}}{\dif y^{s-1}}\Bigg(\frac{h_0(y)}{y}\Bigg)\Bigg]_{y=0}}{eq:AppA_pi_def}
Substituting the expression behind the derivation from Eq.~(\ref{eq:h0_eq}), we get:
\eqarray{\pi_s^{out}	& = & \frac{1}{(s-1)!}\Bigg[\frac{\dif^{s-1}}{\dif y^{s-1}}g_{00}[1,h_1(y)]\Bigg]_{y=0}= \nonumber \\
			& = & \frac{1}{(s-1)!}\Bigg[\frac{\dif^{s-2}}{\dif y^{s-2}}\big[\partial_yg_{00}[1,h_1(y)]h_1'(y)\big]\Bigg]_{y=0}}{eq:AppA_pi_eq1}
This equation can be transformed into an integral by the Cauchy formula:
\eqarray{\pi_s^{out}	& = & \frac{1}{2\pi i(s-1)}\oint\frac{\partial_yg_{00}[1,h_1(y)]}{y^{s-1}}\frac{\dif h_1}{\dif y}\dif y \nonumber \\
			& = & \frac{1}{2\pi i(s-1)}\oint\frac{\partial_yg_{00}[1,h_1]}{y^{s-1}}\dif h_1}{eq:AppA_pi_eq2}
The contour goes around the origin and has an infinitesimal radius, ensuring that it does not enclose poles. In our calculations $y_0=0$. In the last equation we just changed variables. We have to note that when $y\to0$, $h_1$ also converges to zero. This can be easily seen from Eq.~(\ref{eq:h1_eq}).
We can also eliminate $y$ from the argument using Eq.~(\ref{eq:h1_eq}):
\eqarray{\pi_s^{out}	& = & \frac{1}{2\pi i(s-1)}\oint\frac{\partial_yg_{00}[1,h_1]g_{10}[1,h_1]^{s-1}}{h_1^{s-1}}\dif h_1= \nonumber \\
			& = & \frac{\kav}{2\pi i(s-1)}\oint\frac{g_{01}[1,h_1]g_{10}[1,h_1]^{s-1}}{h_1^{s-1}}\dif h_1}{eq:AppA_pi_eq3}
A second application of the Cauchy formula gives us the final form of the out-components:
\eq{\pi_s^{out}=\frac{\kav}{(s-1)!}\Bigg[\frac{\dif^{s-2}}{\dif y^{s-2}}\bigg(g_{01}(1,y)g_{10}(1,y)^{s-1}\bigg)\Bigg]_{y=0}}{eq:AppA_pi_final}

\section{Proof of $\mathbf{\boldsymbol\beta\boldsymbol(u\boldsymbol)\boldsymbol<0}$}
We show that the coefficient
\eq{\beta(u)=\frac{1}{\kav}\frac{\partial_y\chi_{00}(u,1)}{1-\partial_xg_{01}^p(u,1)}}{eq:AppB_beta_def}
that appears in Eq.~(\ref{eq:1pCorr_urho_def}) is negative for any $\kav>k_c$. The numerator has the form of
\eq{\partial_y\chi_{00}(u,1)=\sigma_p^2\sum_{j,k=0}^\infty x^jkm_{jk}}{eq:AppB_dy_khi00_eq1}
where the only criteria for $m_{jk}$ are Eq.~(\ref{eq:1pCorr_mij_crit1})-(\ref{eq:1pCorr_mij_crit2}). They can be satisfied by the following choice \cite{newman03b}:
\eq{m_{jk}=\frac{(p_j-\phi_j)(p_k-\phi_k)}{\big(\kav-\kav_\phi\big)^2}}{eq:AppB_mij_example}
here $\phi_j$ is an arbitrary normalized distribution and $\kav_\phi$ is the corresponding average. With this form of $m_{jk}$, the numerator looks like:
\eq{\partial_y\chi_{00}(u,1)=\sigma_p^2\frac{g_0^p(u)-g_0^\phi(u)}{\kav-\kav_\phi}}{eq:AppB_dy_khi00_eq2}
here $g_0^p(u)$ and $g_0^\phi(u)$ are the corresponding generating functions for the distributions $p_j$ and $\phi_j$. Assume that $\kav>\kav_\phi$ and fix the $\phi_j$ distribution such that $\kav_\phi=k_c$. It can be easily seen that for the ER and exponential networks, $g_0^p(u)<g_0^\phi(u)$ for any $u$.

\begin{figure}
    \centering
	\includegraphics[angle=-90,width=8cm]{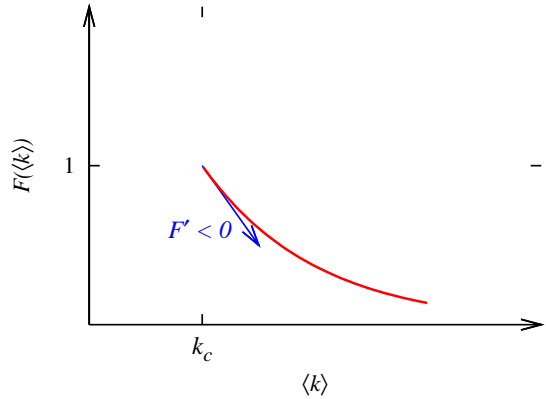}
	\caption{The illustration of $F(\kav)$ for the proof of $\beta(u_0)<0$. Its value at $k_c$ follows directly from the definition and its derivative is examined in the text.}
	\label{fig:beta_negative}
\end{figure}

We only have to show that $\partial_xg_{01}^p(u,1)<1$ above the transition point. For this to see, let us consider the function in question as a function of the average degree:
\eq{F(\kav)=\partial_xg_{01}^p[u(\kav),1]}{eq:AppB_F_def}
and note that $F(k_c)=1$. Furthermore, $F(x)$ is a decreasing function of its argument:
\eq{\frac{\dif F(\kav)}{\dif\kav}=\underbrace{\partial_x^2g_{01}^p(x,1)\big|_{x=u(\kav)}}_{>0}\cdot\underbrace{\frac{\dif u(\kav)}{\dif\kav}}_{<0}}{eq:AppB_dF_dk_eq}
The first term of the right hand side is positive for all $0<u<1$, because it is a sum of positive numbers. In contrast, the second term is negative in the ER and exponential networks, because the size of the giant components increase with the average degree. Thus $F(\kav)<1$ (see Fig.~\ref{fig:beta_negative}).

\bibliography{HierarchyInDirectedRandomNetworks}

\begin{thebibliography}{10}

\bibitem{castellano09}
C.~Castellano, S.~Fortunato, and V.~Loreto.
\newblock Statistical physics of social dynamics.
\newblock {\em Phys. Rev. Lett.}, 81:591--646, 2009.

\bibitem{dunne02}
J.~A. Dunne, R.~J. Williams, and N.~D. Martinez.
\newblock Food-web structure and network theory: {T}he role of connectance and
  size.
\newblock {\em Proceedings of the {N}ational {A}cademy of {S}ciences of the
  {U}nited {S}tates of {A}merica}, 99(20):12917--22, 2002.

\bibitem{jeong00}
H.~Jeong, B.~Tombor, R.~Albert, Z.~N. Oltvai, and A.-L. Barab\'{a}si.
\newblock The large-scale organization of metabolic networks.
\newblock {\em Nature}, 407(6804):651--4, 2000.

\bibitem{negyessy06}
L.~N\'{e}gyessy, T.~Nepusz, L.~Kocsis, and F.~Bazs\'{o}.
\newblock Prediction of the main cortical areas and connections involved in the
  tactile function of the visual cortex by network analysis.
\newblock {\em European Journal of Neuroscience}, 23(7):1919--1930, 2006.

\bibitem{cancho01}
R.~F. Cancho and R.~V. Sol\'{e}.
\newblock The small world of human language.
\newblock {\em Proceedings of the {R}oyal {S}ociety of {L}ondon. {S}eries {B}:
  {B}iological {S}ciences}, 268(1482):2261--2265, 2001.

\bibitem{watts98}
D.~J. Watts and S.~H. Strogatz.
\newblock Collective dynamics of 'small-world' networks.
\newblock {\em Nature}, 393:440--442, 1998.

\bibitem{barabasi02}
R.~Albert and A.-L. Barab\'{a}si.
\newblock Statistical {M}echanics of {C}omplex {N}etworks.
\newblock {\em Phys. Rev. Lett.}, 74:47--97, 2002.

\bibitem{newman10}
M.~E.~J. Newman.
\newblock {\em Networks: {A}n {I}ntroduction}.
\newblock {O}xford {U}niversity {P}ress, 2010.

\bibitem{newman03}
M.~E.~J. Newman.
\newblock The {S}tructure and {F}unction of {C}omplex {N}etworks.
\newblock {\em SIAM Review}, 45(2):167--256, 2003.

\bibitem{pastorsatorras01}
R.~Pastor-Satorras and A.~Vespignani.
\newblock Epidemic spreading in scale-free networks.
\newblock {\em Phys. Rev. Lett.}, 86(14):3200--3203, 2001.

\bibitem{pastorsatorras04}
R.~Pastor-Satorras and A.~Vespignani.
\newblock {\em Evolution and {S}tructure of the {I}nternet}.
\newblock {C}ambridge {U}niversity {P}ress, 2004.

\bibitem{newman01}
M.~E.~J. Newman, S.~H. Strogatz, and D.~J. Watts.
\newblock Random graphs with arbitrary degree distributions and their
  applications.
\newblock {\em Phys. Rev. E}, 64:026118, 2001.

\bibitem{barabasi00}
R.~Albert, H.~Jeong, and A.-L. Barab\'{a}si.
\newblock Error and attack tolerance of complex networks.
\newblock {\em Nature}, 406:378--382, 2000.

\bibitem{bollobas01}
B.~Bollob\'{a}s.
\newblock {\em Random Graphs}.
\newblock Cambridge University Press, 2001.

\bibitem{broder00}
A.~Broder, R.~Kumar, F.~Maghoul, P.~Raghavan, S.~Rajagopalan, R.~Stata,
  A.~Tomkins, and J.~Wiener.
\newblock Graph structure in the web.
\newblock {\em Computer Networks}, 33:309--320, 2000.

\bibitem{dorogovtsev01}
S.~N. Dorogovtsev, J.~F~F. Mendes, and A.~N. Samukhin.
\newblock Giant stronlgy connected component if directed networks.
\newblock {\em Phys. Rev. E}, 64(025101), 2001.

\bibitem{nagy10}
M.~Nagy, Zs. \'{A}kos, D.~Biro, and T.~Vicsek.
\newblock Hierarchical grooup dynamics in pigeon flocks.
\newblock {\em Nature}, 464:890--893, 2010.

\bibitem{fushing11}
H.~Fushing, M.~P. McAssey, B.~Beisner, and B.~McCowan.
\newblock Ranking network of captive rhesus macaque society: {A} sophisticated
  corporative kingdom.
\newblock {\em PLoS ONE}, 6(3):e17817, 2011.

\bibitem{ma04}
H.-W. Ma, J.~Buer, and A.-P. Zeng.
\newblock Hierarchical sructure and modules in the {E}scherichia coli
  transcriptional regulatory network revealed by a new top-down approach.
\newblock {\em BMC Bioinformatics}, 5(1):199, 2004.

\bibitem{huseyn84}
L.~Huseyn and D.~A. Whetten.
\newblock The {C}oncept of {H}orizontal {H}ierarchy and the {O}rganization of
  {I}nterorganizational {N}etworks: a {C}omparative {A}nalysis.
\newblock {\em Social Networks}, 6(1):31--58, 1984.

\bibitem{pumain06}
D.~Pumain, editor.
\newblock {\em Hierarchy in Natural and Social Sciences}.
\newblock Springer: {D}odrecht, the {N}etherlands, 2006.

\bibitem{lane06}
D.~Lane.
\newblock {\em Hierarchy, complexity, society}.
\newblock Springer: Dodrecht, the Netherlands, 2006.

\bibitem{wimberley09}
E.~T. Wimberley.
\newblock {\em Nested ecology. The place of humans in the ecological
  hierarchy}.
\newblock John Hopkins University Press, Maryland, 2009.

\bibitem{carmel02}
L.~Carmel, D.~Haren, and Y.~Koren.
\newblock {\em Drawing {D}irected {G}raphs {U}sing {O}ne-{D}imensional
  {O}ptimization}.
\newblock Springer, 2002.

\bibitem{trusina04}
A.~Trusina, S.~Maslov, P.~Minnhagen, and K.~Sneppen.
\newblock Hierarchy measures in complex networks.
\newblock {\em Phys. Rev. Lett.}, 92(17):178702, 2004.

\bibitem{rowe07}
R.~Rowe, G.~Creamer, S.~Hershkop, and S.~J. Stolfo.
\newblock Automated social hierarchy detection through email network analysis.
\newblock In {\em WebKDD/SNA-KDD ’07: Proceedings of the 9th WebKDD and 1st
  SNA-KDD 2007 workshop on Web mining and social network analysis}, pages
  109--117. ACM, 2007.

\bibitem{memon08}
N.~Memon, H.~L. Larsen, D.~L. Hicks, and N.~Harkiolakis.
\newblock Detecting {H}idden {H}ierarchy in {T}errorist {N}etworks: {S}ome
  {C}ase {S}tudies.
\newblock In {\em Proceedings of the IEEE ISI 2008 PAISI, PACCF, and SOCO
  international workshops on Intelligence and Security Informatics}, pages
  477--489. Springer-Verlag, 2008.

\bibitem{mones12}
E.~Mones, L.~Vicsek, and T.~Vicsek.
\newblock Hierarchical measure for complex networks.
\newblock {\em PLoS ONE}, 7(3):e33799, 2012.

\bibitem{erdos60}
P.~Erd\H{o}s and A.~R\'{e}nyi.
\newblock On the evolution of random graphs.
\newblock {\em Publ. Math. Inst. Hung. Acad. Sci.}, 5:17--60, 1960.

\bibitem{corless96}
R.~M. Corless, G.~H. Gonnet, D.~E.~G. Hare, D.~J Jeffrey, and D.~E. Knuth.
\newblock On the lambert w function.
\newblock {\em Adv. Comput. Math.}, 5:329--359, 1996.

\bibitem{amaral00}
L.~A.~N. Amaral, A.~Scala, M.~Barth\'{e}l\'{e}my, and H.~E. Stanley.
\newblock Classes of small-world networks.
\newblock {\em PNAS}, 97(21):11149--11152, 2000.

\bibitem{goh01}
K.-I. Goh, B.~Kahng, and D.~Kim.
\newblock Universal behavior of load distribution in scale-free networks.
\newblock {\em Phys. Rev. Lett.}, 87(278701), 2001.

\bibitem{chung02}
F.~Chung and L.~Lu.
\newblock Connected component in random graphs with given expected degree
  sequences.
\newblock {\em Annual Combinatorics}, 6:125--145, 2002.

\bibitem{cohen00}
R.~Cohen, K.~Erez, D.~ben{-}Avraham, and S.~Havlin.
\newblock Resilience of the {I}nternet to {R}andom {B}reakdown.
\newblock {\em Phys. Rev. Lett.}, 85(21):4626--4628, 2000.

\bibitem{newman03b}
M.~E.~J. Newman.
\newblock Mixing patterns in networks.
\newblock {\em Phys. Rev. E}, 67(026126), 2003.

\bibitem{newman02}
M.~E.~J. Newman.
\newblock Assortative mixing in networks.
\newblock {\em Phys. Rev. Lett.}, 89(208701), 2002.

\bibitem{boguna03}
M.~Bogun\'{a}, R.~Pastor-Satorras, and A.~Vespignani.
\newblock {A}bsence of {E}pidemic {T}hreshold in {S}cale-{F}ree {N}etworks with
  {D}egree {C}orrelations.
\newblock {\em Phys. Rev. Lett.}, 90:028701, 2003.

\bibitem{martinez91}
N.~Martinez.
\newblock Artifacts or attributes? {E}ffects of resolution on the {L}ittle
  {R}ock {L}ake food web.
\newblock {\em Ecological Monographs}, 61:367--392, 1991.

\bibitem{circuits}
\texttt{http://courses.engr.illinois.edu/ece543/iscas89.html}.

\bibitem{leskovec10}
J.~Leskovec, D.~Huttenlocher, and J.~Kleinberg.
\newblock Signed networks in social media.
\newblock In {\em Proceedings of the 28th international conference on Human
  factors in computing systems}, pages 1361--1370. ACM, 2010.

\bibitem{vanduijn03}
M.~A.~J. Van~Duijn, M.~Huisman, F.~N. Stokman, F.~W. Wasseur, and E.~P.~H.
  Zeggelink.
\newblock Evolution of sociology freshmen into a friendship network.
\newblock {\em Journal of Mathematical Sociology}, 27:153--191, 2003.

\bibitem{milo04}
R.~Milo, S.~Itzkovitz, N.~Kashtan, R.~Levitt, S.~Shen-Orr, I.~Ayzenshtat,
  M.~Sheffer, and U.~Alon.
\newblock Superfamilies of evolved and designed networks.
\newblock {\em Science}, 303(5663):1538--42, 2004.

\bibitem{balaji06}
S.~Balaji, M.~M. Babu, L.~M. Iyer, N.~M. Luscombe, and L.~Aravind.
\newblock Comprehensive analysis of combinatorial regulation using the
  transcriptional regulatory network of yeast.
\newblock {\em Journal of Molecular Biology}, 360(1):213--27, 2006.

\bibitem{milo02}
R.~Milo, S.~Shen-Orr, S.~Itzkovitz, N.~Kashtan, D.~Chklovskii, and U.~Alon.
\newblock Network motifs: simple building blocks of complex networks.
\newblock {\em Science}, 298(5594):824--7, 2002.

\end{thebibliography}

\end{document}